\pdfoutput=1
\documentclass[12pt]{article}

\usepackage{epsfig}
\usepackage{cite}

\usepackage{amsmath}  
\usepackage{amssymb}
\usepackage{euscript}

\newcommand{\Keu}{\EuScript{K}}

\begin{document}

\begin{titlepage}

\begin{flushright}
\bf IFJPAN-IV-2008-6\\
\end{flushright}

\vspace{1mm}
\begin{center}
  {\LARGE\bf%
    Markovian MC simulation of QCD evolution at NLO level with minimum $k_T$
$^{\star}$
}
\end{center}
\vspace{2mm}

\begin{center}
{\large\bf P.~Stok\l{}osa$^{a}$, 
W.~P\l{}aczek$^b$,
{\rm and} \large\bf M.~Skrzypek$^{a}$}
\end{center}
\vspace{2mm}

\begin{center}
{\em $^a$Institute of Nuclear Physics IFJ-PAN,\\
  ul.\ Radzikowskiego 152, 31-342 Cracow, Poland.}\\ \vspace{2mm}
{\em $^b$Marian Smoluchowski Institute of Physics, Jagiellonian University,\\
   ul.\ Reymonta 4, 30-059 Cracow, Poland.}\\ \vspace{2mm}
\end{center}

\vspace{3mm}
\begin{abstract}
We present two Monte Carlo algorithms of the Markovian type which solve the
modified QCD evolution equations at the NLO level. The modifications with
respect to the standard DGLAP evolution concern the argument of the strong
coupling constant $\alpha_S$. We analyze the $z$-dependent argument and then the
$k_T$-dependent one. The evolution time variable is
identified with the rapidity. The two algorithms are tested to
the $0.05$\% precision level. We find that the NLO corrections in the evolution of parton momentum distributions with $k_T$-dependent coupling constant are of the order of 10 to 20\%, and in a small $x$ region even up to 30\%, with respect to the LO contributions.
\end{abstract}

\vspace{3mm}
\begin{center}
\end{center}

\vspace{3mm}
\begin{flushleft}
\bf IFJPAN-IV-2008-6\\
\end{flushleft}

\vspace{5mm}
\footnoterule
\noindent
{\footnotesize
$^{\star}$The project is partly supported by the EU grant MTKD-CT-2004-510126,
  realized in the partnership with the CERN Physics Department and by the
  Polish Ministry of Science and Information Society Technologies
  grant No 620/E-77/6.PR UE/DIE 188/2005-2008.
}

\end{titlepage}

\section{Introduction}
With the first LHC results coming soon the precision of the
QCD calculations becomes an important issue. As usuall, one finds two
approaches to the calculations: fixed-order calculations and resumed to all orders Monte Carlo (MC) parton showers. The
fixed-order results are impressive. Let us give a few examples of such calculations: 
the splitting
functions, calculated up to NNLO level \cite{Moch:2004pa,Vogt:2004mw} or various differential and semi-inclusive distributions, calculated in the NNLO and even NNNLO approximation see e.g.\
\cite{Anastasiou:2005qj,Grazzini:2008tf,GehrmannDeRidder:2007hr}.
On the other hand, the MC parton shower approach, indispensable in describing
complicated experimental signal selection procedures, does not achieve such a high
precision. In fact, the {\em complete} NLO level has not been reached yet by any
of the available MC parton shower codes. In most of the cases these codes are based on some form of an {\em improved} LO approximation, see e.g.\ 
\cite{Sjostrand:2007gs,Sjostrand:2000wi,Gieseke:2006ga,Corcella:2000bw,Lonnblad:1992tz}.
Another approach is based on combining the NLO matrix element with the LO parton shower
\cite{Frixione:2002ik,Nason:2004rx}.

Some time ago some of us started the MC parton shower project with the
ambitious goal of reaching the NLO precision level. The project is based on
the DGLAP evolution equation \cite{DGLAP} at the NLO level 
\cite{Floratos:1977au,Floratos:1978ny,Curci:1980uw,Furmanski:1980cm}. We began by creating two different algorithms, and then constructing two MC programs, that solve the evolution
equation and simulate the one-hemisphere collinear parton shower. 
{\em First algorithm}, called
{\tt EvolFMC} (Markovian Monte Carlo), is based on the principle of a Markovian
process. It generates the DGLAP-type evolution of Parton Distribution Functions (PDFs) up to the NLO level including
both gluons and quarks. In Ref.\ \cite{Jadach:2003bu} we have shown for the first time that with
the aid of modern CPU such a MC code can solve the LO evolution equations with the
precision below $10^{-3}$, i.e.\ comparable or even better than the other numerical
methods. This study has been extended to the NLO evolution in Ref.\ \cite{GolecBiernat:2006xw}.
The {\em second algorithm} called CMC is an entirely new type of an algorithm, an example 
of a wider class that we named "Constrained Markovian Monte Carlo". CMC is designed to
solve the Markovian evolution {\it with superimposed constraint} on
both $x$ and flavor type of the outgoing parton
\cite{Jadach:2005bf,Jadach:2007qa}. Such an evolution with
predefined end-point, mandatory for the simulation of the resonant processes,
is an alternative for the commonly used workaround solution of this problem
based on the so-called "backward evolution" \cite{Sjostrand:1985xi}.
The ultimate goal is
to combine two initial-state non-collinear NLO constrained evolutions (in two hemispheres) together with the hard process into one NLO parton shower for the Drell-Yan-type processes, either in full agreement with the DGLAP evolution or including effects of a modified-DGLAP type. 

The {\tt EvolFMC} code includes also the modified DGLAP
evolutions in which the argument of the coupling constant is replaced by more
complicated functions of $x$, $x'$ and $Q^2$
\cite{Amati:1980ch, Brodsky:1982gc, Wong:1995kn,Sterman:1986aj, Ermolaev:2001zs}. 
In particular,
the use of $k_T$ as the argument is known to effectively resum some of the
higher order soft corrections \cite{Amati:1980ch,Sterman:1986aj,Ermolaev:2008dq}. For this reason it is the preferred
choice of most of the MC parton shower codes.
The $k_T$ is used as an argument also by the CCFM evolution equation \cite{CCFM}. This equation is designed to effectively interpolate between the DGLAP evolution in the large $x$ region and the BFKL-type behavior in the low $x$ region. The CCFM equation uses, apart from the modified arguments of coupling constant and angular ordering, also the "non-Sudakov form factor", important in the small $x$ region.

In the case of the developed by us
{\tt EvolFMC} code the option of modified argument of the coupling constant has been so far
implemented only in the LO approximation \cite{GolecBiernat:2007pu,Skrzypek:2007zz}. 
This is a significant limitation since both the NLO
kernels and the $k_T$-dependent coupling constant are important for the final
precision of the parton shower. In the present paper we will fill in this gap
and present the NLO extension of the $k_T$-dependent {\tt EvolFMC} code. 
In this scheme the evolution variable is understood to be the rapidity of emitted partons, ensuring the angular ordering.
We will discuss also another choice of the argument: $Q(1-z)$. It is based on the variant of the CCFM equation, called the "one-loop" CCFM \cite{Marchesini:1990zy,Kwiecinski:2002}.
Of course, $Q(1-z)$ is only one of a few $z$-dependent functions that have been used in the literature \cite{Brodsky:1982gc, Wong:1995kn, Roberts:1999gb}.

The primary role of the presented here Markovian NLO code would be to serve as a powerfull testing device capable of reproducing independently any of the above evolution types. In particular, the emulation of the CCFM equation would be possible, if the collinear evolution is supplied with the transverse momenta and the "non-Sudakov form factor" is added. Another area of application of such a Markovian MC would be in performing fits of the PDFs to the $F_2$ data. We have demonstrated recently \cite{Stoklosa:2007zz} that using MC codes for the purpose of fitting is feasible with the modern computer power. Finally, for the simulation of the final-state parton shower, the Markovian algorithm without constraints is directly applicable.

Let us digress here that the standard evolution kernels of the collinear factorisation, either LO or NLO, are used in the normal integrated form, i.e.\ as functions of the $x$ and $t$ variables only. This is of course the well-established and extremely successful DGLAP approach. Nonetheless, for more exclusive observables at the LHC, it will be necessary to include the effects of transverse momenta beyond the collinear approximation. Some attempts in this direction have already been made, for example by the above mentioned MC$@$NLO \cite{Frixione:2002ik,Nason:2004rx}, the GR$@$PPA project \cite{Tsuno:2006cu,Kurihara:2002ne} or the CCFM-based \cite{CCFM} CASCADE project \cite{Jung:2000hk}.
 
Another interesting approach is
based on the idea of "exclusive" DGLAP kernels in which the internal degrees of freedom are {\em not} integrated out analytically, but instead simulated (i.e.\ integrated) by the MC parton shower itself \cite{stjinprep}.

Finally, let us comment on yet another class of MC algorithms solving the evolution equations -- the veto algorithms \cite{Seymour:1994df,Sjostrand:2006za,GolecBiernat:2007xv}. They are based on a Markovian evolution with additional internal pseudo-emission loop. Thanks to this feature it is enough to compute an approximate Sudakov form factor, instead of the exact one, for each of the generated events and the algorithm becomes faster. On the negative side, the average multiplicity of the generated partons is much higher. We did not use the veto scheme in this work. The main reason for this is that, as discussed earlier, we consider Markovian-type algorithms primarily as a testing device for the constrained Markovian algorithms. In these latter cases the veto algorithm does not apply and the Sudakov form factors must be calculated. For that reason we preferred to have the Sudakov form factors implemented (and mutually tested) in both approaches.

This paper is organized as follows. In the next Section we introduce all
necessary notation, briefly present the general formalism of the Markovian MC
solutions of the evolution equations and define two particular evolution
schemes that we are interested in.
In the following two Sections we describe in detail the MC
algorithms that solve the above two evolutions by means of the collinear Markovian
parton shower. We present two algorithms, main and auxiliary (mostly for
tests), for each of the schemes. 
As the simpler ones we discuss in detail the $Q(1-z)$-type algorithms, whereas in the description of the $k_T$-dependent ones we focus mostly on the modifications with respect to the $Q(1-z)$-type algorithms.
In Section 5 we present numerical results:
first we discuss the choice of the counter term and input parameters, then we
briefly describe a variety of technical tests that we performed in order to establish the
technical precision of the {\tt EvolFMC} code, and finally we compare and discuss various
types of evolution. The Summary Section concludes the paper.

\section{General framework}
\label{sect:general}

In this paper we will follow the general formulation of the evolution equation
and its solutions in terms of the Markovian MC presented in Ref.\
\cite{GolecBiernat:2007xv}. In order to establish the notation let us recall here basic
definitions and formulas. For more complete presentation we refer the reader
to the Ref.\ \cite{GolecBiernat:2007xv}. We write the evolution equation in the compact
matrix notation 
\begin{equation}
\label{eq:op-evoleq}
  \partial_t {\bf D}(t) = {\bf K}(t) \; {\bf D}(t)
\end{equation}
and in the explicit representation
\begin{equation}
\label{eq:genevoleq}
  \partial_t D_f(t,x)
 = \sum_{f'} \int_x^1 dw\; \Keu_{ff'}(t,x,w) D_{f'}(t,w),
\end{equation}
where $D_{f}(t,w)$ denotes the parton density function of the parton $f$
and $\Keu_{ff'}(t,x,w)$ denotes the generalized evolution kernel. Note that
contrary to the DGLAP case, it depends on two variables, $x$ and $w$, instead of
their ratio $z=x/w$ only. The kernel is then decomposed into real and virtual
parts
\begin{equation}
\label{eq:evkernel}
\begin{split}
&\Keu_{ff'}(t,x,w)= \Keu^V_{ff'}(t,x,w)+\Keu^R_{ff'}(t,x,w),\quad
 \Keu^V_{ff'}(t,x,w)= -\delta_{ff'}\delta_{x=w}\Keu^v_{ff}(t,x).
\end{split}
\end{equation}
The Markovian solution of eq.\ (\ref{eq:op-evoleq}) is conveniently
expressed with the help of the operator $\bar{\bf E}$ defined as
\begin{equation}
\label{eq:momsumrule}
  \bar{\bf E}\; {\bf D}(t)
  \equiv \int_0^1 dx \sum_{f} x\; D_f(t,x), 
\quad\hbox{i.e.}\quad \bar E_f(x)\equiv x.
\end{equation}
This operator in a formal way shows that we use an "unconstrained" evolution --
all values of $x$ and $f$ will be generated and that the evolution will be done
for momentum distributions $x\bf D(t)$. The solution of eq.\
(\ref{eq:op-evoleq}) and the master formula for the Markovian MC is then
\begin{equation}
\label{eq:evol5}
\begin{split}
\bar{\bf E}{\bf D}(t)=
\int_{t_{0}}^{t} dt_1    \bigg(
\int_{t_{1}}^{t} dt_2    \bigg[&
\int_{t_{2}}^{t} dt_3    \bigg\{ \dots
\\
\dots
\int_{t_{N-1}}^{t} dt_{N}\bigg\{
&  \bar{\bf E}{\bf K}^R(t_{N})
   {\bf G}_{{\bf K}^V}(t_{N},t_{N-1})
  +\bar{\bf E}{\bf G}_{{\bf K}^V}(t,t_{N-1})\delta_{t_{N}=t}
\bigg\}
\\~~~~~~~~~~~~~ \vdots
\\ \times
&  {\bf K}^R(t_{2})
   {\bf G}_{{\bf K}^V}(t_{2},t_{1})
  +\bar{\bf E}{\bf G}_{{\bf K}^V}(t,t_{1})\delta_{t_2=t}
\bigg]
\\ \times
&
 {\bf K}^R(t_{1})
 {\bf G}_{{\bf K}^V}(t_{1},t_{0})
+\bar{\bf E}{\bf G}_{{\bf K}^V}(t,t_{0})\delta_{t_1=t}
\bigg)
{\bf D}(t_0),
\end{split}
\end{equation}
where the diagonal matrix ${\bf G}_{{\bf K}^V}$ is the solution of the evolution
equation with the virtual kernel $\Keu^V_{ff'}(t,x,w)$ parametrized in terms of the Sudakov form factor $\Phi_f(t,t'|x)$:
\begin{equation}
\label{eq:phidef}
\{{\bf G}_{{\bf K}^V}(t,t')\}_{ff'}(x,w)
  =\delta_{ff'}\delta_{x=w}\; e^{-\Phi_f(t,t'|w)},\qquad
\Phi_f(t,t'|x)=\int_{t'}^{t} dt''\; \Keu^v_{ff}(t'',x).
\end{equation}
The above formula (\ref{eq:evol5}) follows from the iteration of the momentum conservation principle
\begin{equation}
\int\limits_{t_{i-1}}^{t} dt_i\;
\Big\{
  \bar{\bf E}{\bf K}^R(t_{i})
   {\bf G}_{{\bf K}^V}(t_{i},t_{i-1})
  +\bar{\bf E}{\bf G}_{{\bf K}^V}(t,t_{i-1})\delta_{t_{i}=t}
\Big\} =
\bar{\bf E}.
\label{crucial}
\end{equation}
Eq.\ (\ref{crucial}) defines also the probabilities of the single step forward
in
$t_i,f_i$ and $x_i$ variables:
\begin{equation}
\label{eq:omega}
\begin{split}
1&= \frac{1}{x_{i-1}}
\int\limits_{t_{i-1}}^{t} dt_i\;
\Big\{
  \bar{\bf E}{\bf K}^R(t_{i})
   {\bf G}_{{\bf K}^V}(t_{i},t_{i-1})
  +\bar{\bf E}{\bf G}_{{\bf K}^V}(t,t_{i-1})\delta_{t_{i}=t}
\Big\}_{f_{i-1}}(x_{i-1})
\\
&=
  e^{-\Phi_{f_{i-1}}(t,t_{i-1}|x_{i-1})}
+\int\limits^1_{e^{-\Phi_{f_{i-1}}(t,t_{i-1}|x_{i-1})}}
 d\left(e^{-\Phi_{f_{i-1}}(t_{i},t_{i-1}|x_{i-1})}\right)
\\& ~~~~\times
 \left[
 \sum_{f_i}
 \frac{\partial_{t_i}\Phi_{f_i f_{i-1}}(t_{i},t_{i-1}|x_{i-1})}%
      {\partial_{t_i}\Phi_{f_{i-1}}(t_{i},t_{i-1}|x_{i-1})}
 \int dx_i\;
 \frac{1}{\partial_{t_i}\Phi_{f_i f_{i-1}}(t_{i},t_{i-1}|x_{i-1})}
 \frac{x_i}{x_{i-1}}
 \Keu^R_{f_if_{i-1}}(t_{i},x_i,x_{i-1})
 \right],
\end{split}
\end{equation}
where the virtual Sudakov form factor is expressed in terms of the real emission part of the evolution kernel
\begin{equation}
\label{eq:phireal}
\begin{split}
\Phi_{f_{i-1}} (t_i, t_{i-1} |x_{i-1})
  =&\int\limits^{t_i}_{t_{i-1}} d t\; {\Keu^v_{f_{i-1}f_{i-1}} (t, x_{i-1})} 
\\
  =&\sum_{f_i} 
   \int\limits^{t_i}_{t_{i-1}} d t 
   \int\limits_0^{x_{i-1}} \frac{d x_i}{x_{i-1}} x_i\;
      \Keu^R_{f_i f_{i-1}}(t,x_i,x_{i-1})
  =\sum_{f_i} \Phi_{f_i f_{i-1}} (t_i, t_{i-1} |x_{i-1}).
\end{split}
\end{equation}
In the case of a weighted algorithm with the global correcting weight, one uses
the simplified kernel 
$\Keu^R_{f_if_{i-1}}(t_{i},x_i,x_{i-1})\to
 \bar{\Keu}^R_{f_if_{i-1}}(t_{i},x_i,x_{i-1})$.
This simplification is compensated by the global weight
\begin{equation}
\label{wtcorr}
w^{(n)}
= e^{\bar\Phi_{f_{n}}(t,t_{n} |x_{n})-\Phi_{f_{n}}(t,t_{n} |x_{n})}
  \left(
  \prod_{i=1}^n 
\frac{      \Keu^R_{f_if_{i-1}}(t_{i},x_i,x_{i-1})}%
           {\bar{\Keu}^R_{f_if_{i-1}}(t_{i},x_i,x_{i-1})} 
  e^{\bar\Phi_{f_{i-1}}(t_{i},t_{i-1} |x_{i-1})
          -\Phi_{f_{i-1}}(t_{i},t_{i-1}|x_{i-1})}
  \right),
\end{equation}
where $n$ denotes the number of emissions generated within a given MC event,
and 
the form factor $\bar\Phi_{f_{i-1}}(t_{i},t_{i-1}|x_{i-1})$ is constructed from
$\bar{\Keu}^R_{f_if_{i-1}}(t_{i},x_i,x_{i-1})$ in complete analogy to eq.\
(\ref{eq:phireal}).
The last quantity to be defined here is the exact shape of the kernels,
including the definition of the argument of the coupling constant in terms of the $t$ and $x$ variables. Following Ref.\ \cite{GolecBiernat:2007xv}
we will discuss two schemes of modified-DGLAP type, denoted in Ref.\ \cite{GolecBiernat:2007xv} as
(B') and (C'). The novelty with respect to Ref.\ \cite{GolecBiernat:2007xv} is that we will
perform the calculation and construct the Markovian MC code at the NLO level,
whereas in Ref.\ \cite{GolecBiernat:2007xv} only the LO case has been discussed.
To be specific, the schemes are defined as
\begin{equation}
\label{kernBare}
\begin{split}
&x\Keu^{R(B')}_{f'f}(t,x,w)=
\\
&=      \biggl(
  \frac{\alpha_{NLO}(\ln(1-z)+t)}{2\pi}2zP_{f'f}^{R(0)}(z)
+ \Bigl(\frac{\alpha_{NLO}(\ln(1-z)+t)}{2\pi}\Bigr)^2 2zP_{f'f}^{R(1)B'}(z)
 \biggr) \theta_{1-z>\lambda e^{-t}},
\\
&x\Keu^{R(C')}_{f'f}(t,x,w)=
\\
&=      \biggl(
  \frac{\alpha_{NLO}(\ln(w-x)+t)}{2\pi}2zP_{f'f}^{R(0)}(z)
+ \Bigl(\frac{\alpha_{NLO}(\ln(w-x)+t)}{2\pi}\Bigr)^2 2zP_{f'f}^{R(1)C'}(z)
 \biggr) \theta_{w-x>\lambda e^{-t}},
\end{split}
\end{equation}
where $z=x/w$ and $\lambda$ is the cut-off on the argument of the coupling
constant. Note that $\lambda$ is not an infinitesimal IR cut-off, as used in
the DGLAP case. On the contrary, $\lambda$ is finite, typically of the order of
a few GeV.
Note also that the factor 2 in front of the kernels is due to our
definition of the evolution time, which in the DGLAP case is chosen as $t=\ln
Q$ rather than $t=\ln Q^2$.

The NLO parts of the kernels, $P_{f'f}^{R(1)B'}(z)$ and $P_{f'f}^{R(1)C'}(z)$, 
consist of the universal part $P_{f'f}^{R(1)}(z)$ \cite{Curci:1980uw,Furmanski:1980cm} 
and the
evolution scheme dependent counter terms 
$\Delta P_{f'f}^{R(1)B'}(z)$ and $\Delta P_{f'f}^{R(1)C'}(z)$
\begin{align}
\begin{split}
\label{non-universal}
P_{f'f}^{R(1)B'}(z) =& P_{f'f}^{R(1)}(z)+\Delta P_{f'f}^{R(1)B'}(z),
\\
P_{f'f}^{R(1)C'}(z) =& P_{f'f}^{R(1)}(z)+\Delta P_{f'f}^{R(1)C'}(z).
\end{split}
\end{align}
These counter terms are necessary to remove the double counting introduced by
the shift of the arguments of the coupling constants. There is some freedom in
defining these counter terms. The algorithms presented here work for any
(reasonable) choice of the counter terms. For further details on the choice we
used in this work we refer the reader to Section \ref{sect:double}.

We will frequently be using also the representation of the universal
parts of the kernels based on
their structure in the $z$-variable
\begin{align}
zP_{f'f}^{R(0)}(z) =&\,\frac{1}{1-z}\delta_{f'f} A_{ff}^{(0)}
                    +F_{f'f}^{(0)}(z),
\notag
\\
zP_{f'f}^{R(1)}(z) =&\,\frac{1}{1-z}\delta_{f'f} A_{ff}^{(1)}(z)
                    +F_{f'f}^{(1)}(z).
\end{align}
The functions $P_{f'f}^{R}(z)$,
$A_{ff}^{}(z)$ and
$F_{f'f}(z)$, written in the notation used here, are collected in Ref.\
\cite{GolecBiernat:2006xw}.

The coupling constant at the NLO level has the standard
form
\begin{align}
\label{alpha}
    \alpha_{LO}(t) &= \frac{2\pi}{\beta_0(t-\ln\Lambda_0)},
\notag \\
    \alpha_{NLO}(t) &= \alpha_{LO}(t)\left(
          1-\alpha_{LO}(t) \
              \frac{\beta_1 \ln(2t-2\ln\Lambda_0)}{4\pi\beta_0}
                                        \right).
\end{align}

We close this general introduction with a brief explanation why we state that
the argument $\ln(x_{i-1}-x_i)+t_i$ can be regarded as $k^T_i$ of the emitted real parton with the four momentum $k_i$. 
It follows immediately from the kinematical mapping of the evolution variables into four momenta, provided that the evolution time is identified with the rapidity of the emitted parton
and the $x$-variable, as usual, with the light-cone plus component of the virtual parton 
\begin{align}
t_i = \ln(2E_h) -\frac{1}{2}\ln\frac{k_i^+}{k_i^-},\;\;\;
k_i^+ =2E_h (x_{i-1}-x_i)
\end{align}
where $E_h$ is an arbitrary reference energy of the incoming hadron. 
As a consequence of the maslessness of $k_i$ we obtain
\begin{align}
k^T_{i} = \sqrt{k^+ k^-} =e^{t_i}(x_{i-1}-x_i).
\end{align}

Let us now proceed with the description of the novel NLO algorithms for
the two schemes.
\section{Markovian algorithm for scheme (B')}
In this section we will present two schemes of solving the evolution B' in a
Markovian way at the NLO level. We begin with the efficient one and then we
present the other scheme, devised mostly for the testing purposes.
\subsection{Main algorithm}
The main algorithm is based on the simplified LO DGLAP kernel in which the
coupling constant is used in the NLO approximation. Specifically we follow
the eq.\ (3.21) of Ref.\ \cite{GolecBiernat:2007pu}, see also \cite{Placzek:2007xb},
and we extend it to the NLO case 
\begin{align}
\label{simplebprim}
x\bar\Keu^{R(B')}_{f'f}(t,x,w)
\equiv&
\frac{\alpha_{NLO}(t+\ln(1-z))}{2\pi}
2z\bar P^{R(B')}_{f'f}(z)\theta_{1-z>\lambda e^{-t}},
\\
z\bar P^{R(B')}_{f'f}(z)=&
\frac{1}{1-z}
    (\delta_{f'f} A_{ff}^{(0)}+\max_z F^{(0)}_{f'f}(z) +M_{f'f}).
\end{align}
We remind the reader that $z=x/w$.
Note that
the singular factor $1/(1-z)$ is artificially introduced into the $F$-part in order
to achieve analytical integrability of the formula, see Ref.\ \cite{GolecBiernat:2007pu} for
more discussion.
The constant $M_{f'f}$ is defined as
\begin{align}
M_{f'f} = \left\{ \begin{array}{c}
0,\;\;\;\;{\rm if\;} P^{R(0)}_{f'f}(z)\neq 0,
\\
\eta,\;\;\;\;{\rm if\;} P^{R(0)}_{f'f}(z) = 0,
\end{array} \right.
\end{align}
where $\eta$ is a dummy technical parameter. The
reason behind introduction of $M_{f'f}$ is very simple -- we want to remove all
zeroes in the LO transition matrix, because some of them might become non-zero
at the NLO level and cause infinite weights. The constant
$\eta$ is therefore added to all kernels that are zero at the LO level. Of
course this is purely technical, dummy, operation, later on corrected by means
of a proper rejection weight.

The corresponding Sudakov form factor, necessary to generate the time $t_i$, is then
defined as
\begin{align}
 	\bar{\Phi}_f^{B'}(t_i,t_{i-1}|x_{i-1})=&
     \int_{t_{i-1}}^{t_i}dt\int_0^1 d\Bigl(\frac{x}{w}\Bigr) \sum_{f'}  
   x\bar\Keu^{R(B')}_{f'f}(t,x,w)
\notag
\\
=&  \frac{1}{\pi}\int_{t_{i-1}}^{t_i} dt
    \int^{t-t_\lambda}_{0}du\,
\alpha_{NLO}(t-u)
	(A_{ff}^{(0)}+\sum_{f'} (\max_z F^{(0)}_{f'f}(z)+M_{f'f})) 
\notag
\\=& \int_{t_{i-1}}^{t_i} dt
    \Bigl(\rho(t,t-t_\lambda)-\rho(t,0)\Bigr) 
    \Bigl(A_{ff}^{(0)}+\sum_{f'} \max_z (F^{(0)}_{f'f}(z)+M_{f'f})\Bigr) 
\notag
\\
	 =& (\bar\zeta(t_i)-\bar\zeta(t_{i-1}) )
         (A_{ff}^{(0)}+\sum_{f'} (\max_z F^{(0)}_{f'f}(z)+M_{f'f})),
\label{phibarb}
\end{align}
where
\begin{align}
  \label{rho}
   \rho(t,u)
  &= \frac{1}{\pi} \int du\, \alpha_{NLO}(t-u)
\notag
\\ &=
	\frac{1}{\pi} \int 
{du}\frac{2}{ \beta_{0}  G(t,u)   }\left( 1-\frac{1}{2}\,{\frac {\beta_1\ln 
	(2 G(t,u)) }{{
	\beta_0}^{2}G(t,u) }} \right)
\notag \\
&=-\left(\frac {2}{\beta_0} +{\frac {\beta_1}{{\beta_0}^{3}
	G(t,u)}}\right)\ln (2G(t,u)) - {\frac {\beta_1}{{\beta_0}^{3}G(t,u) }}
           \notag
         \\
          G(t,u) &=  t  -u - \ln \Lambda_0 ,\qquad  u= -\ln
(1-z),\qquad t_\lambda=\ln\lambda,
\end{align}
and
\begin{align}
	\bar{\zeta}(t)&=
          a_{00} \left( \ln  \left( t-\ln \Lambda_0\right)  \right) ^{2}+
\left( a_{10}t+a_{11} \right) \ln  \left( t-\ln \Lambda_0\right) +a_{20}t,
\\
	\notag a_{00}&=\frac{1}{2}\,{\frac {\beta_1}{{\beta_0}^{3}}},
\\
	\notag a_{10}&=\frac{2}{\beta_0},
\\
	\notag a_{11}&=-{\frac {2\,{\beta_0}^{2}\ln \Lambda_0 -\beta_1
	\ln 2 -\beta_1}{{\beta_0}^{3}}},
\\
 \notag a_{20}&=-\frac{\beta_1+2\beta_0^2(t_\lambda-\ln \Lambda_0)}
 {\beta_0^3(t_\lambda-\ln \Lambda_0)}\ln(t_\lambda-\ln \Lambda_0)
 -\frac{2\beta_0^2(t_\lambda-\ln \Lambda_0)+\beta_1(1+\ln 2)}{\beta_0^3(t_\lambda-\ln \Lambda_0)}	.
\end{align}
\\
Let us note that the coefficients $a_{ij}$ need to be calculated only once during
initialization of the algorithm. The procedure of inverting the function
$\zeta(t_i)$, necessary to generate the time $t_i$, has to be done numerically.

In the next step we generate the flavor index $f_i$ based on the probability
\begin{align}
\label{flavindex}
    p_{f_i}&=\frac{\partial_{t_i} 
    \bar{\Phi}_{f_{i}f_{i-1}}^{B'}(t_i,t_{i-1}|x_{i-1})}
              {\partial_{t_i} 
    \bar{\Phi}^{B'}_{f_{i-1}}(t_i,t_{i-1}|x_{i-1})}
   =\frac{\delta_{f_i f_{i-1}}A_{f_{i-1}f_{i-1}}^{(0)}  
                 + \max_z F^{(0)}_{f_i f_{i-1}}(z)+M_{f_i f_{i-1}}}
       {A_{f_{i-1}f_{i-1}}^{(0)}
    +  \sum_{f_i}(\max_z F^{(0)}_{f_i f_{i-1}}(z)+M_{f_i f_{i-1}})},
    \;\;\; \sum_{f_i} p_{f_i} =1,
\end{align}
where
\begin{align}
     {\partial_{t_i} 
     \bar{\Phi}^{B'}_{f_if_{i-1}}(t_i,t_{i-1}|x_{i-1})}=&
     \int_0^1 d\Bigl(\frac{x}{w}\Bigr) 
   x\bar\Keu^{R(B')}_{f_{i}f_{i-1}}(t_i,x,w)
\notag
\\
	=& 
    \Bigl(\rho(t_i,t_i-t_\lambda)-\rho(t_i,0)\Bigr) 
    \Bigl(\delta_{f_i f_{i-1}}A_{f_{i-1}f_{i-1}}^{(0)}
            +\max_z F^{(0)}_{f_{i}f_{i-1}}(z)+M_{f_i f_{i-1}}\Bigr)
\end{align}
and
\begin{align}
     {\partial_{t_i} 
    \bar{\Phi}^{B'}_{f_{i-1}}(t_i,t_{i-1}|x_{i-1})}=&
   \sum_{f_i}   {\partial_{t_i} 
     \bar{\Phi}^{B'}_{f_if_{i-1}}(t_i,t_{i-1}|x_{i-1})}.
\end{align}

As the last variable we generate $z_i$. 
The normalized density distribution $dz_ip^{B'}(z_i)$ is given by 
\begin{align}
\label{zdistr}
 dz_ip^{B'}(z_i)=& dz_i  \frac{\alpha_{NLO}(t_i+ \ln(1-z_i))}{\pi}
         \frac{\Theta_{1-z_i>\lambda e^{-t_i}}}{1-z_i}
    \biggl(
    \int_0^1 dz \frac{\alpha_{NLO}(t_i+ \ln(1-z))}{\pi}
         \frac{\Theta_{1-z>\lambda e^{-t_i}}}{1-z}
    \biggr)^{-1} 
\notag
\\ 
   =& du_i \frac{\alpha_{NLO}(t_i-u_i)}{\pi}
         \Theta_{t_i-t_\lambda>u_i>0}
	\Bigl(\rho(t_i,t_i-t_\lambda) -\rho(t_i,0)\Bigr)^{-1} 
\notag
\\ 
   =& du_i \partial_{u_i}\rho(t_i,u_i) 
         \Theta_{t_i-t_\lambda>u_i>0}
	\Bigl(\rho(t_i,t_i-t_\lambda) -\rho(t_i,0)\Bigr)^{-1}.
\end{align}
In the actual generation of $z_i$ we use the
method of inverse cumulative. The $\rho(t_i,u_i)$
function has to be inverted numerically with respect to the $u_i$ variable.

The last part of the algorithm to be discussed here is the correcting weight,
defined in eq.\ (\ref{wtcorr}), compensating the simplification of the kernel
done in eq.\ (\ref{simplebprim}). The most complicated part of this weight is
related to the exact Sudakov form factor. It has the form of the
double integral. Numerical evaluation of such a double integral would
significantly slow down the algorithm. Therefore it is essential to perform at
least one of the integrations analytically. In the following we will show how
this can be done in the NLO case.

\begin{figure}[!ht]
  \centering
    \epsfig{file=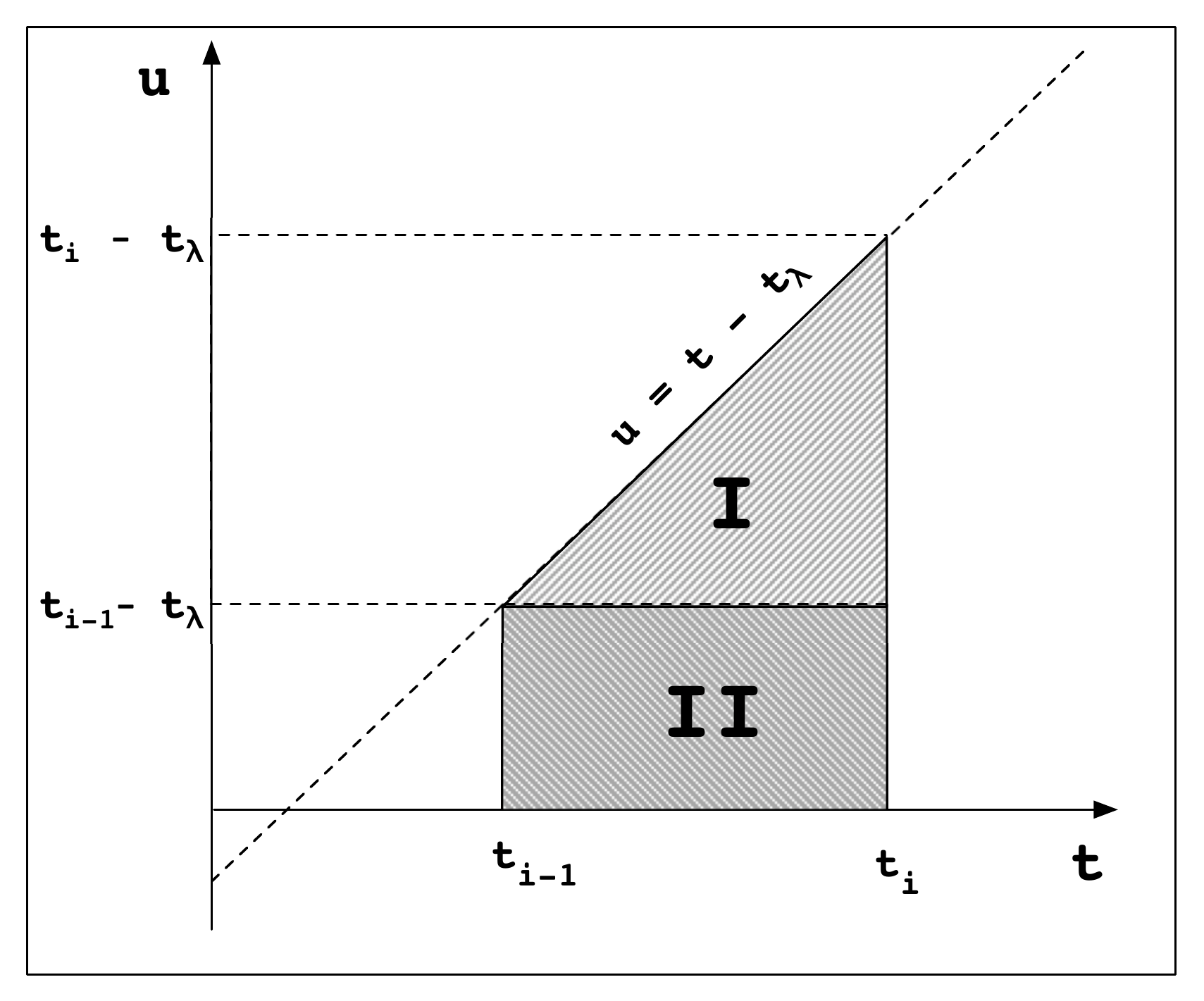,width=80mm}
  \caption{\sf
   The integration domain in the $tu$ space for the case B'. Regions I and II
correspond to
two integrals in eq.\ (\ref{zmiana_b}).
    }
  \label{fig:tuspaceb}
\end{figure}

Let us define the full Sudakov form factor $\Phi_f^{B'}(t_i,t_{i-1}|x_{i-1})$
of the B' evolution
\begin{align}
 \Phi_f^{B'}(t_i,t_{i-1}|x_{i-1})=&
     \int_{t_{i-1}}^{t_i}dt\int_0^1 d\Bigl(\frac{x}{x_{i-1}}\Bigr) \sum_{f'}  
   x\Keu^{R(B')}_{f'f}(t,x,x_{i-1})
\notag
\notag
\\
=&
\int_{t_{i-1}}^{t_i}dt\int_0^1 dz \sum_{f'}
     \biggl(
   \frac{\alpha_S(t+\ln(1-z))}{2\pi}2zP_{f'f}^{R(0)}(z)
\notag
\\
& + \Bigl(\frac{\alpha_S(t+\ln(1-z))}{2\pi}\Bigr)^2 2zP_{f'f}^{R(1)B'}(z)
  \biggr) \theta_{1-z>\lambda e^{-t}},
\end{align}
and let us write down the desired virtual component of the weight
\begin{align}
\label{delta}
\Delta_{f}^{B'}(t_i,t_{i-1}|x_{i-1})=&
   \Phi_f^{B'}(t_i,t_{i-1}|x_{i-1}) -\bar{\Phi}_f^{B'}(t_i,t_{i-1}|x_{i-1})
\notag
\\
=&
\int_{t_i}^{t_{i-1}}dt\int_0^1 dz  \theta_{1-z>\lambda e^{-t}}\sum_{f'}
     \biggl(
   \frac{\alpha_S(t+\ln(1-z))}{2\pi}2z
   \bigl(P_{f'f}^{R(0)}(z) -\bar P_{f'f}^{R(0)}(z)\bigr)
\notag
\\ &
 + \Bigl(\frac{\alpha_S(t+\ln(1-z))}{2\pi}\Bigr)^2 2zP_{f'f}^{R(1)B'}(z)
  \biggr).
\end{align}
The integral over $dz$ for general form of the kernel cannot be performed
analytically even at the LO level (cf.\ Ref.\ \cite{GolecBiernat:2007pu}). However, as in the LO
case of Ref.\ \cite{GolecBiernat:2007pu}, the $dt$ integral can be done analytically also for the
NLO case. The calculation looks as follows. We introduce the usual variable
$u=-\ln(1-z)$ and then change order of integration. The resulting integral can
be expressed as a sum of two integrals over the two regions of
the $tu$ space shown in Fig. \ref{fig:tuspaceb}%
\footnote{The similar change of the integration order was also exploited
by the authors of HERWIG MC~\cite{Corcella:2000bw,Marchesini:1988cf}.}:
\\
\begin{align}
\int_{t_{i-1}}^{t_i}dt\int_0^1 dz \theta_{1-z>\lambda e^{-t}}
=&
\int_{t_{i-1}}^{t_i}dt \int_{0}^{t-t_\lambda}du\, e^{-u}
\notag
\\
=&
\int_{t_{i-1}-t_\lambda}^{t_i-t_\lambda}du e^{-u}
\int_{u+t_\lambda}^{t_i} dt 
    +
\int_0^{t_{i-1}-t_\lambda}du e^{-u}
\int_{t_{i-1}}^{t_i} dt.
\label{zmiana_b}
\end{align}
The $dt$ integral, which depends on the coupling constant only, can be done
analytically with the help of the integrals
\begin{align}
	\frac{1}{\pi}J_1(t,u)=&  \frac{1}{\pi}\int dt \alpha_{NLO}(t-u) 
      =-\rho(t,u)
\label{j1}
\end{align}
and
\begin{align}
\notag  \frac{1}{\pi^2}J_2(t,u) &=  \frac{1}{\pi^2}\int dt
   \alpha_{NLO}^2(t-u) 
\\
\notag 
&=
\int dt {\frac { \left(
     -2\,{\beta_0}^{2}(t-u-\ln \Lambda_0) 
     +\beta_1\ln  2\left( t-u-\ln \Lambda_0
	\right) \right) ^2}
                {\beta_0^{6} \left( t-u-\ln \Lambda_0  \right) ^{4}}}
\\
\notag
&=
 \frac{a_{00}}{G^3(t,u)} \ln^2 (2G(t,u))
 +\biggl( \frac{a_{10}}{G^3(t,u)}
         +\frac{a_{11}}{G^2(t,u)}\biggr)\ln (2G(t,u)) 
\\
&\;\;\;
 +\frac{a_{20}}{G^3(t,u)}
 +\frac{a_{30}}{G^2(t,u)}
 +\frac{a_{40}}{G(t,u)},	
\label{g2}
\end{align}
where
\begin{align}
\notag 
a_{00}&=-\frac{1}{3}\frac{\beta_1^2}{\beta_0^6},\;\;\;
a_{10}=-\frac{2}{9}\frac{\beta_1^2}{\beta_0^6},\;\;\;
a_{11}=\frac{4 \beta_1}{\beta_0^4},\;\;\;
a_{20}=-\frac{2}{27}\frac{\beta_1^2}{\beta_0^6},\;\;\;
a_{30}=\frac{ \beta_1}{\beta_0^4},\;\;\;
a_{40}=-\frac{4}{\beta_0^2}.
\notag
\end{align}
Collecting together the above results we can rewrite eq.\ (\ref{delta}) as
\begin{align}
\label{delta2}
\Delta_{f}^{B'}&(t_i,t_{i-1}|x_{i-1})=
\notag
\\
=&
\int_{t_{i-1}-t_\lambda}^{t_i-t_\lambda}du e^{-u} 2z
\sum_{f'} \biggl( 
   \frac{(J_1(t_i,u)-J_1(u+t_\lambda,u))}{2\pi}
   \bigl(P_{f'f}^{R(0)}(z) -\bar P_{f'f}^{R(B')}(z)\bigr)
\notag
\\ &+
   \frac{(J_2(t_i,u)-J_2(u+t_\lambda,u))}{4\pi^2}P_{f'f}^{R(1)B'}(z)
          \biggr)
\notag
\\ &+
\int_0^{t_{i-1}-t_\lambda}du e^{-u} 2z
\sum_{f'} \biggl( 
   \frac{(J_1(t_i,u)-J_1(t_{i-1},u))}{2\pi}
   \bigl(P_{f'f}^{R(0)}(z) -\bar P_{f'f}^{R(B')}(z)\bigr)
\notag
\\ &+
   \frac{(J_2(t_i,u)-J_2(t_{i-1},u))}{4\pi^2} P_{f'f}^{R(1)B'}(z)
          \biggr)
\end{align}
where $z=1-e^{-u}$.
Note the cancellation of the leading singularity, $A_{ff}^{(0)}/(1-z)$, in the
above formula due to
\begin{align}
\sum_{f'} z
\Bigl( P_{f'f}^{R(0)}(z) -\bar P_{f'f}^{R(B')}(z) \Bigr)
=
\sum_{f'} 
\Biggl( F_{f'f}^{(0)}(z)  - \frac{1}{1-z}\biggl( \max_z F_{f'f}^{(0)}(z)+M_{f'f} \biggr)
          \Biggr).
\end{align}
\\
Combining together the real and virtual components we arrive at
the final formula for the global weight of eq.\ (\ref{wtcorr}) adopted to the
case of the scheme B'
\begin{align}
\label{wtcorrb}
w^{(n)(B')}
= e^{-\Delta_{f_{n}}^{B'}(t,t_{n}|x_{n})}
  \prod_{i=1}^n 
  \left(
\frac{      \Keu^{R(B')}_{f_if_{i-1}}(t_{i},x_i,x_{i-1})}%
           {\bar{\Keu}^{R(B')}_{f_if_{i-1}}(t_{i},x_i,x_{i-1})} 
  e^{-\Delta_{f_{i-1}}^{B'}(t_{i},t_{i-1}|x_{i-1})}
  \right).
\end{align}

\subsection{Auxiliary algorithm}
\label{auxB}
The main algorithm described in previous section is a fairly complicated one,
especially due to the presence of numerical integrations and numerical
inversions of
various functions. Therefore, it is obligatory to devise a way of testing it
down to the sub-per-mille precision level. We have not found any
non-Monte-Carlo program that solves the modified-DGLAP-type evolutions at the NLO
level and therefore we
constructed another, independent Monte Carlo algorithm for the purpose of cross-checks. The
algorithm is less efficient but at the same time it is simpler.

The algorithm is closely based on the LO algorithm described in Ref.\ \cite{GolecBiernat:2007pu}.
The entire NLO correction is introduced as a weight.
We will only briefly sketch it here and we refer the interested
reader to Ref.\ \cite{GolecBiernat:2007pu} for details.

The algorithm is based on the simplified kernel of the form
\begin{align}
\label{simplebprimLO}
x\bar\Keu^{R(B')(2)}_{f'f}(t,x,w) 
\equiv& 
\frac{\alpha_{LO}(t+\ln(1-z))}{2\pi}
2z\bar P^{R(B')}_{f'f}(z)\theta_{1-z>\lambda e^{-t}}.
\end{align}
As compared to simplified kernel (\ref{simplebprim}) of the previous algorithm
in this one the coupling constant is taken in the LO approximation.

In complete analogy to eq.\ (\ref{phibarb}) the simplified Sudakov form factor
reads
\begin{align}
 	\bar{\Phi}_f^{B'(2)}(t_i,t_{i-1}|x_{i-1})=&
     \int_{t_{i-1}}^{t_i}dt\int_0^1 d\Bigl(\frac{x}{x_{i-1}}\Bigr) \sum_{f'} 
   x\bar\Keu^{R(B')(2)}_{f'f}(t,x,x_{i-1})
\notag
\\
=&  \frac{1}{\pi}\int_{t_{i-1}}^{t_i} dt\int^{t-t_\lambda}_{0}du
\alpha_{LO}(t-u) (A_{ff}^{(0)}+\sum_{f'} (\max_z F^{(0)}_{f'f}(z)+M_{f'f})) 
\notag
\\
	=& \int_{t_{i-1}}^{t_i} dt
    \Bigl(\rho_{LO}(t,t-t_\lambda)-\rho_{LO}(t,0)\Bigr) 
    \Bigl(A_{ff}^{(0)}+\sum_{f'} (\max_z F^{(0)}_{f'f}(z)+M_{f'f})\Bigr)
    \notag
\\ 
	 =& (\bar\zeta_{LO}(t_i)-\bar\zeta_{LO}(t_{i-1}) )
         (A_{ff}^{(0)}+\sum_{f'} (\max_z F^{(0)}_{f'f}(z)+M_{f'f})),
\label{phibarbintLO}
\end{align}
where
\begin{align}
  \label{rhoLO}
   \rho_{LO}(t,u)
  &= \frac{1}{\pi} \int du\, \alpha_{LO}(t-u)
=- \frac{2}{\beta_0} \ln (2G(t,u)).
\end{align}
The function $\bar\zeta_{LO}(t)$ reads
\begin{align}
 	\bar\zeta_{LO}(t_i)=\frac{2}{\beta_0}(t_i-\ln \Lambda_0)(\ln(t_i-\ln
\Lambda_0)-1) 
\end{align}
In order to generate $t_i$ the function $\bar\zeta_{LO}(t_i)$ must be inverted.
This inversion can be done analytically.
However, in the actual Monte Carlo, for the purpose of tests we
have implemented also the numerical inversion procedure, similar as in the case
of the first algorithm (there is no visible computing time overhead related to this
numerical inversion).

The generation of the flavor index $f_i$ is based on the probability identical
to eq.\ (\ref{flavindex})
\begin{align}
\label{flavindex2}
    p_{f_i}^{(2)}&=\frac{\partial_{t_i}
\bar{\Phi}_{f_{i}f_{i-1}}^{B'(2)}(t_i,t_{i-1})}
              {\partial_{t_i} \bar{\Phi}^{B'(2)}_{f_{i-1}}(t_i,t_{i-1})}
   =\frac{\delta_{f_i f_{i-1}}A_{f_{i-1}f_{i-1}}^{(0)}  
                           + \max_z F^{(0)}_{f_i f_{i-1}}(z)+M_{f_i f_{i-1}}}
       {A_{f_{i-1}f_{i-1}}^{(0)}
    +  \sum_{f_i}\biggl(\max_z F^{(0)}_{f_i f_{i-1}}(z)
   +M_{f_i f_{i-1}}\biggr)}
   \equiv p_{f_i}.
\end{align}
It is the case because the coupling constants cancel in both eqs.\
(\ref{flavindex}) and (\ref{flavindex2}).
The generation of the $z$-variable is based on the LO version of eq.\
(\ref{zdistr})
\begin{align}
\label{zdistrLO}
 dz_ip^{B'(2)}(z_i)=& dz_i  \frac{\alpha_{LO}(t_i+ \ln(1-z_i))}{\pi}
         \frac{\Theta_{1-z_i>\lambda e^{-t_i}}}{1-z_i}
    \biggl(
    \int_0^1 dz \frac{\alpha_{LO}(t_i+ \ln(1-z))}{\pi}
         \frac{\Theta_{1-z>\lambda e^{-t_i}}}{1-z}
    \biggr)^{-1} 
\notag
\\ 
   =& du_i \partial_{u_i}\rho_{LO}(t_i,u_i) 
         \Theta_{t_i-t_\lambda>u_i>0}
	\Bigl(\rho_{LO}(t_i,t_i-t_\lambda) -\rho_{LO}(t_i,0)\Bigr)^{-1}.
\end{align}
Contrary to the previous algorithm, now the function $\rho_{LO}(t,u)$ can be
inverted
analytically. However, for the purpose of testing various components of the
program, we have implemented also an option of the numerical inversion.

As indicated earlier, the novelty of this algorithm, i.e.\ the NLO effect, is
hidden in the global weight. The most complicated part of the weight is the
virtual component built out of the Sudakov form factors. For the calculation of
the exact form factor we can use the results derived for the previous algorithm, and
the whole virtual part of the weight follows from eq.\ (\ref{delta})
\begin{align}
\label{deltaLO}
\Delta_{f}^{B'(2)}&(t_i,t_{i-1}|x_{i-1})=
   \Phi_f^{B'}(t_i,t_{i-1}|x_{i-1}) 
  -\bar{\Phi}_f^{B'(2)}(t_i,t_{i-1}|x_{i-1})
\notag
\\
=&
\int_{t_i}^{t_{i-1}}dt\int_0^1 dz  \theta_{1-z>\lambda e^{-t}}\sum_{f'}
     \biggl(
    \frac{\alpha_{NLO}(t+\ln(1-z))}{2\pi}2z
   P_{f'f}^{R(0)}(z)
\notag
\\ &
 + \Bigl(\frac{\alpha_{NLO}(t+\ln(1-z))}{2\pi}\Bigr)^2 2zP_{f'f}^{R(1)B'}(z)
 - \frac{\alpha_{LO}(t+\ln(1-z))}{2\pi}2z
   \bar P_{f'f}^{R(0)}(z)
  \biggr),
\end{align}
with the final result
\begin{align}
\label{delta2LO}
\Delta_{f}^{B'(2)}&(t_i,t_{i-1}|x_{i-1})=
\notag
\\
=&
\int_{t_{i-1}-t_\lambda}^{t_i-t_\lambda}du e^{-u} 2z
\sum_{f'} \biggl( 
      -
   \frac{(J_1^{LO}(t_i,u)-J_1^{LO}(u+t_\lambda,u))}{2\pi}
   \bar P_{f'f}^{R(B')}(z)
\notag
\\ &    +
   \frac{(J_1(t_i,u)-J_1(u+t_\lambda,u))}{2\pi}
   P_{f'f}^{R(0)}(z) 
   +
   \frac{(J_2(t_i,u)-J_2(u+t_\lambda,u))}{4\pi^2}P_{f'f}^{R(1)B'}(z)
          \biggr)
\notag
\\ &+
\int_0^{t_{i-1}-t_\lambda}du e^{-u} 2z
\sum_{f'} \biggl( 
     -
   \frac{(J_1^{LO}(t_i,u)-J_1^{LO}(t_{i-1},u))}{2\pi}
   \bar P_{f'f}^{R(B')}(z)
\notag
\\ &+
   \frac{(J_1(t_i,u)-J_1(t_{i-1},u))}{2\pi}
   P_{f'f}^{R(0)}(z) 
    +
   \frac{(J_2(t_i,u)-J_2(t_{i-1},u))}{4\pi^2} P_{f'f}^{R(1)B'}(z)
          \biggr),
\end{align}
where $z=1-e^{-u}$ and
\begin{align}
	\frac{1}{\pi}J_1^{LO}(t,u)=&  \frac{1}{\pi}\int dt \alpha_{LO}(t-u) 
      =-\rho_{LO}(t,u).
\end{align}

We conclude this section by presenting the complete formula for the global
weight in this algorithm:
\begin{equation}
\label{wtcorrbLO}
w^{(n)(B')(2)}
= e^{-\Delta_{f_{n}}^{B'(2)}(t,t_{n}|x_{n})}
  \prod_{i=1}^n 
  \left(
\frac{      \Keu^{R(B')}_{f_if_{i-1}}(t_{i},x_i,x_{i-1})}%
           {\bar{\Keu}^{R(B')(2)}_{f_if_{i-1}}(t_{i},x_i,x_{i-1})} 
  e^{-\Delta_{f_{i-1}}^{B'(2)}(t_{i},t_{i-1}|x_{i-1})}
  \right).
\end{equation}
\section{Markovian algorithm for scheme (C')}
Having completed the presentation of the algorithms for the scheme B' we
proceed now to the most important scheme C'. As discussed in Ref.\ \cite{GolecBiernat:2007xv},
it can be interpreted as evolution in the rapidity variable with $k_T$ as the argument of the coupling constant and it is of great physical
importance. The NLO algorithms for the scheme C' are quite similar to the
algorithms for the B' scheme. Therefore in the following sections we will
skip some of the details and concentrate on the differences with respect to the
scheme B'. As before we
begin with the efficient one and then we
present the other algorithm used for tests.
\subsection{Main algorithm}
The main algorithm is based on the simplified kernel similar to the one used 
in the scheme B' (i.e.\ the LO kernel with the NLO coupling constant):
\begin{align}
\label{simplecprim}
x\bar\Keu^{R(C')}_{f'f}(t,x,w) 
\equiv& 
\frac{\alpha_{NLO}(t+\ln w+\ln(1-z))}{2\pi}
2z\bar P^{R(C')}_{f'f}(z)\theta_{(1-z)w>\lambda e^{-t}},
\\
z\bar P^{R(C')}_{f'f}(z)\equiv &z\bar P^{R(B')}_{f'f}(z)=
 \frac{1}{1-z}
     (\delta_{f'f} A_{f'f}^{(0)}+\max_z F^{(0)}_{f'f}(z)+M_{f'f}).
\end{align}
Note that now the kernel depends truly on both the $x$ and
$w$ variables (through the $\theta$-function). In the B' case it depended
only on the ratio $z=x/w$.
The simplified Sudakov form factor, necessary to generate time $t_i$, is then
defined as
\begin{align}
 	\bar{\Phi}_f^{C'}&(t_i,t_{i-1}|x_{i-1})=
     \int_{t_{i-1}}^{t_i}dt\int_0^1 d\left(\frac{x}{x_{i-1}}\right) \sum_{f'}  
    x\bar\Keu^{R(C')}_{f'f}(t,x,x_{i-1}) 
\notag
\\
=&  \frac{1}{\pi}\int_{t_{i-1}}^{t_i} dt\int^{t+\ln x_{i-1}-t_\lambda}_{0}du\,
\alpha_{NLO}(t+\ln x_{i-1} - u)
	\Bigl(A_{ff}^{(0)}+\sum_{f'} (\max_z F^{(0)}_{f'f}(z)+M_{f'f})\Bigr) 
\notag
\\
	=& \int_{t_{i-1}}^{t_i} dt
    \Bigl(\rho(t+\ln x_{i-1},t+\ln x_{i-1}-t_\lambda)
             -\rho(t+\ln x_{i-1},0)\Bigr) 
    \Bigl(A_{ff}^{(0)}+\sum_{f'} (\max_z F^{(0)}_{f'f}(z)+M_{f'f})\Bigr) 
\notag
\\
   =& \Bigl(\bar\zeta(t_i+\ln x_{i-1})-\bar\zeta(t_{i-1}+\ln x_{i-1}) \Bigr)
    \Bigl(A_{ff}^{(0)}+\sum_{f'} (\max_z F^{(0)}_{f'f}(z)+M_{f'f})\Bigr).
\label{phibarcint}
\end{align}
The functions $\rho$ and $\bar\zeta$ are the same as in the algorithm B'. The
only difference is in the shifted $t$-argument: 
$t \to t+\ln x_{i-1}$ or, equivalently, shifted
integration/generation limits
$t_{i(i-1)} \to t_{i(i-1)}+\ln x_{i-1}$. 

The generation of the flavor index $f_i$ is done, identically as in the
previous algorithms, by means of the probabilities $p_{f_i}$ defined in eqs.\ 
(\ref{flavindex}) or (\ref{flavindex2}).

As the last variable we generate $z_i$ using the
integrand of the density
function $dz_ip^{C'}(z_i)$ given by the formula
\begin{align}
\label{zdistrc}
 dz_ip^{C'}(z_i)=& dz_i  \frac{\alpha_{NLO}(t_i+ \ln x_{i-1}+\ln(1-z_i))}{\pi}
         \frac{\Theta_{(1-z_i)x_{i-1}>\lambda e^{-t_i}}}{1-z_i}
\notag
\\ & \times
    \biggl(
    \int_0^1 dz \frac{\alpha_{NLO}(t_i+ \ln x_{i-1}+ \ln(1-z))}{\pi}
         \frac{\Theta_{(1-z)x_{i-1}>\lambda e^{-t_i}}}{1-z}
    \biggr)^{-1} 
\notag
\\ 
   =& du_i \frac{\alpha_{NLO}(t_i+ \ln x_{i-1}-u_i)}{\pi}
         \Theta_{t_i+ \ln x_{i-1}-t_\lambda>u_i>0}
\notag
\\ & \times
	\Bigl(\rho(t_i+ \ln x_{i-1},t_i+ \ln x_{i-1}-t_\lambda)
                 -\rho(t_i+ \ln x_{i-1},0)\Bigr)^{-1} 
\notag
\\ 
   =& du_i \partial_{u_i}\rho(t_i+ \ln x_{i-1},u_i) 
         \Theta_{t_i+ \ln x_{i-1}-t_\lambda>u_i>0}
\notag
\\ & \times
	\Bigl(\rho(t_i+ \ln x_{i-1},t_i+ \ln x_{i-1}-t_\lambda)
                  -\rho(t_i+ \ln x_{i-1},0)\Bigr)^{-1}.
\end{align}
As in the case of $t_i$-variable we observe that the whole difference with
respect to the case B' is in the shift 
$t_{i} \to t_{i}+\ln x_{i-1}$. 

\begin{figure}[!ht]
  \centering
    \epsfig{file=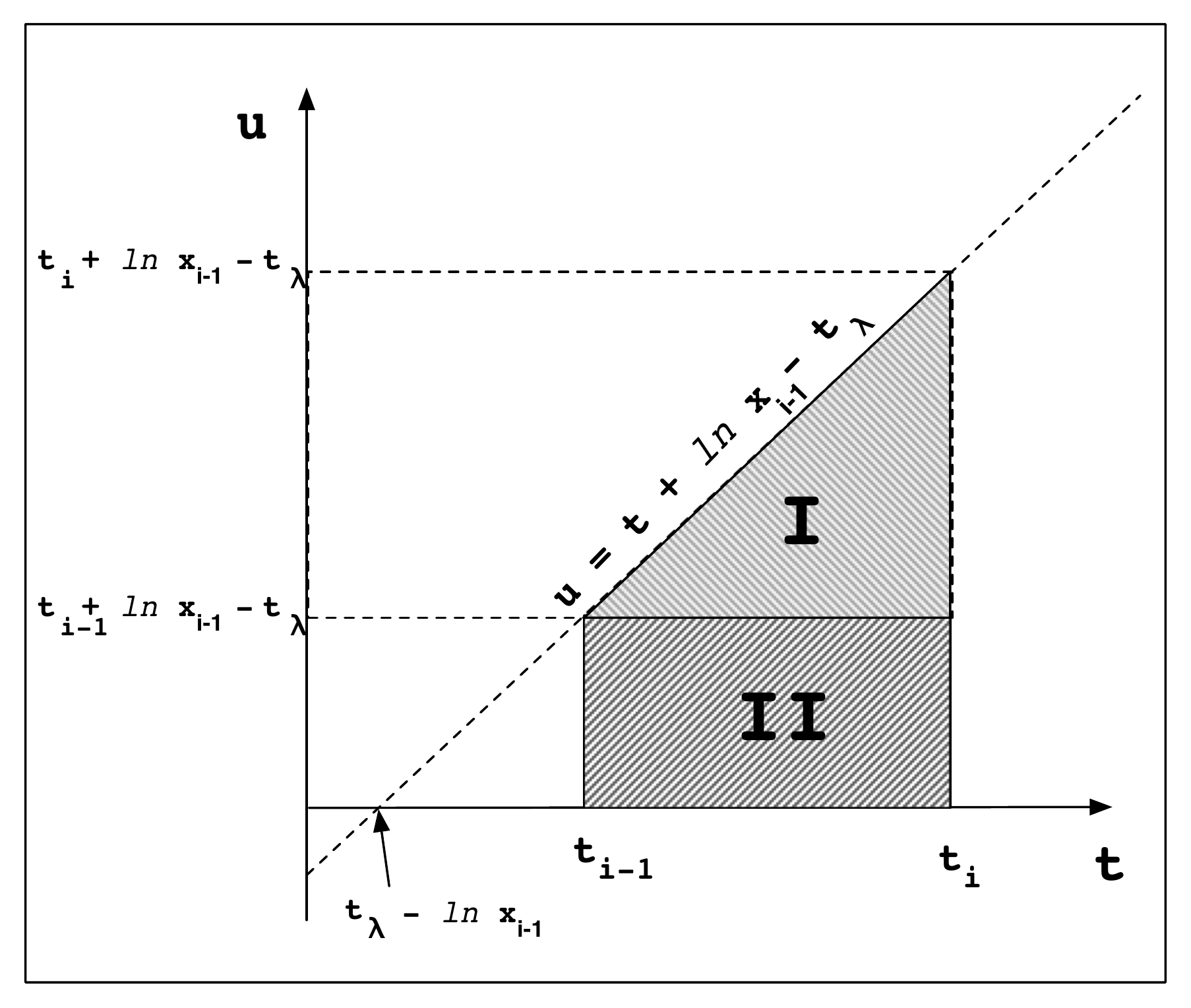,width=78.2mm}
    \epsfig{file=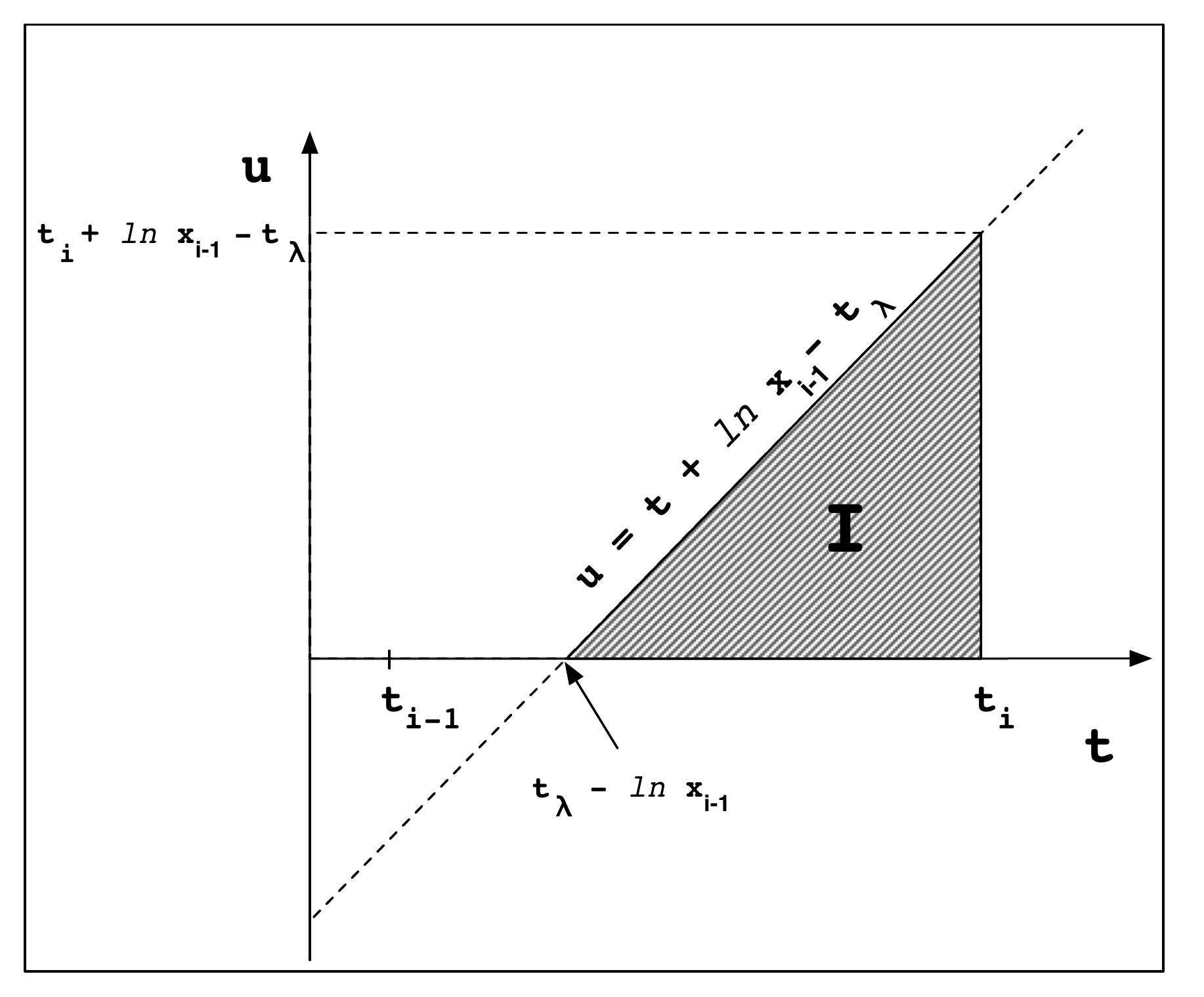,width=80mm}
  \caption{\sf
   The integration domain in the $tu$ space for the case C'. 
Regions I (triangle) and II (rectangle)
correspond to two integrals in eq.\ (\ref{zmiana_c}), respectively.
\underline{Left frame}: the case of ${t_\lambda-\ln x_{i-1}<t_{i-1}}$.
\underline{Right frame}: the case of ${t_\lambda-\ln x_{i-1}\geqslant t_{i-1}}$.
          }
  \label{fig:tuspacec_full}
\end{figure}

Let us define now the full Sudakov form factor
$\Phi_f^{(C')}(t_i,t_{i-1}|x_{i-1})$
\begin{align}
 &\Phi_f^{(C')}(t_i,t_{i-1}|x_{i-1})=
     \int_{t_{i-1}}^{t_i}dt\int_0^1 d\Bigl(\frac{x}{x_{i-1}}\Bigr) \sum_{f'}  
   x\Keu^{R(C')}_{f'f}(t,x,x_{i-1})
\end{align}
As in the case B' we are able to perform analytically only one of the
integrations in $\Phi_f^{(C')}$. To this end we rearrange order of integrations
and decompose the integrals as follows
\begin{align}
\int_{t_{i-1}}^{t_i}&dt\int_0^1 dz \theta_{(1-z)x_{i-1}>\lambda e^{-t}}
=
\int_{t_{i-1}}^{t_i}dt \int_{0}^{t+\ln x_{i-1}-t_\lambda}du\, e^{-u}
\notag
\\
=&\theta_{t_\lambda-\ln x_{i-1}<t_{i-1}} \Biggl(
\int_{t_{i-1}+\ln x_{i-1}-t_\lambda}^{t_i+\ln x_{i-1}-t_\lambda}du e^{-u}
\int_{u-\ln x_{i-1} + t_\lambda}^{t_i} dt
+\int_0^{t_{i-1}+\ln x_{i-1}-t_\lambda}du e^{-u}
\int_{t_{i-1}}^{t_i} dt \Biggr)
\notag
\\
&+
\theta_{t_\lambda-\ln x_{i-1} \geqslant t_{i-1}}
\int_{t_{i-1}+\ln x_{i-1}-t_\lambda}^{t_i+\ln x_{i-1}-t_\lambda}du e^{-u}
\int_{u-\ln x_{i-1} + t_\lambda}^{t_i} dt  
\\
=&
\int_{t_{i-1}+\ln x_{i-1}-t_\lambda}^{t_i+\ln x_{i-1}-t_\lambda}du e^{-u}
\int_{u-\ln x_{i-1} + t_\lambda}^{t_i} dt 
+\theta_{t_\lambda-\ln x_{i-1}<t_{i-1}} 
\int_0^{t_{i-1}+\ln x_{i-1}-t_\lambda}du e^{-u}
\int_{t_{i-1}}^{t_i} dt. 
\label{zmiana_c}
\end{align}
Note that the decomposition changes when $t_\lambda-\ln x_{i-1}$ becomes
smaller/greater than $t_{i-1}$. These two cases are depicted in Fig.
\ref{fig:tuspacec_full}.
With the help of the rearrangement (\ref{zmiana_c}) we can calculate the virtual component
\begin{align}
\Delta_{f}^{C'}&(t_i,t_{i-1}|x_{i-1})=
   \Phi_f^{C'}(t_i,t_{i-1}|x_{i-1}) -\bar{\Phi}_f^{C'}(t_i,t_{i-1}|x_{i-1})
\notag
\\
=&
\int_{t_i}^{t_{i-1}}dt\int_0^1 dz  \theta_{(1-z)x_{i-1}>\lambda
e^{-t}}\sum_{f'}
     \biggl(
   \frac{\alpha_S(t+\ln x_{i-1}+\ln(1-z))}{2\pi}2z
   \bigl(P_{f'f}^{R(0)}(z) -\bar P_{f'f}^{R(0)}(z)\bigr)
\notag
\\ &
 + \Bigl(\frac{\alpha_S(t+\ln x_{i-1}+\ln(1-z))}{2\pi}\Bigr)^2
2zP_{f'f}^{R(1)C'}(z)
  \biggr)
\notag
\\
=&
\int_{t_{i-1}+\ln  x_{i-1}-t_\lambda}^{t_i+\ln x_{i-1}-t_\lambda}du e^{-u} 2z
\sum_{f'}\Biggl(  
   \frac{J_1(t_i+\ln x_{i-1},u)-J_1(u+t_\lambda,u)}{2\pi}
   \bigl(P_{f'f}^{R(0)}(z) -\bar P_{f'f}^{R(C')}(z)\bigr)
\notag
\\ &+
   \frac{J_2(t_i+\ln x_{i-1},u)-J_2(u+t_\lambda,u)}{4\pi^2}P_{f'f}^{R(1)C'}(z)\Biggr)  
\notag
\\ &+
 \theta_{t_{\lambda}-\ln  x_{i-1}<t_{i-1}}
 \int_0^{t_{i-1}+\ln x_{i-1}-t_\lambda}du e^{-u} 2z
\sum_{f'} 
   \notag
   \\
   &  \biggl( 
   \frac{J_1(t_i+\ln x_{i-1},u)-J_1(t_{i-1}+\ln x_{i-1},u)}{2\pi}
   \bigl(P_{f'f}^{R(0)}(z) -\bar P_{f'f}^{R(C')}(z)\bigr)
\notag
\\ &+
   \frac{J_2(t_i+\ln x_{i-1},u)-J_2(t_{i-1}+\ln x_{i-1},u)}{4\pi^2} P_{f'f}^{R(1)C'}(z)
          \biggr) ,\;\;\;\; z=1-e^{-u}.
\label{deltac}
\end{align}
Finally, the
formula for the global weight, corresponding to eq.\ (\ref{wtcorrb}), in
the case C' reads
\begin{equation}
\label{wtcorrc}
w^{(n)(C')}
= e^{-\Delta_{f_{n}}^{C'}(t,t_{n}|x_{n})}
  \prod_{i=1}^n 
  \left(
\frac{      \Keu^{R(C')}_{f_if_{i-1}}(t_{i},x_i,x_{i-1})}%
           {\bar{\Keu}^{R(C')}_{f_if_{i-1}}(t_{i},x_i,x_{i-1})} 
  e^{-\Delta_{f_{i-1}}^{C'}(t_{i},t_{i-1}|x_{i-1})}
  \right).
\end{equation}
\subsection{Auxiliary algorithm}
\label{auxC}
Similarly to the case B', the second algorithm is based on the version
of the simplified kernel (\ref{simplecprim}) in which the coupling constant
is taken in the LO approximation:
\begin{align}
\label{simplecprimLO}
x\bar\Keu^{R(C')(2)}_{f'f}(t,x,w) 
\equiv& 
\frac{\alpha_{LO}(t+\ln w+\ln(1-z))}{2\pi}
2z\bar P^{R(C')}_{f'f}(z)\theta_{(1-z)w>\lambda e^{-t}}.
\end{align}
As in the case B',
in analogy to eq.\ (\ref{phibarcint}), the Sudakov form factor becomes
\begin{align}
 	\bar{\Phi}_f^{C'(2)}&(t_i,t_{i-1}|x_{i-1})=
     \int_{t_{i-1}}^{t_i}dt\int_0^1 d\Bigl(\frac{x}{x_{i-1}}\Bigr) \sum_{f'} 
   x\bar\Keu^{R(C')(2)}_{f'f}(t,x,x_{i-1})
\notag
\\
=&  \frac{1}{\pi}\int_{t_{i-1}}^{t_i} dt
                 \int^{t+\ln x_{i-1}-t_\lambda}_{0}du
         \alpha_{LO}(t-u) 
  \biggl(A_{ff}^{(0)}+\sum_{f'} \Bigl(\max_z
F^{(0)}_{f'f}(z)+M_{f'f}\Bigr)\biggr) 
\notag
\\
	=& \int_{t_{i-1}}^{t_i} dt
    \Bigl( \rho_{LO}(t+\ln x_{i-1},t+\ln x_{i-1}-t_\lambda)
          -\rho_{LO}(t+\ln x_{i-1},0)
    \Bigr) 
\notag
\\
   & \biggl(A_{ff}^{(0)}+\sum_{f'} 
         \Bigl(\max_z F^{(0)}_{f'f}(z)+M_{f'f}\Bigr)\biggr)
   \notag
\\ 
	 =& \Bigl( \bar\zeta_{LO}(t_i+\ln x_{i-1})
                  -\bar\zeta_{LO}(t_{i-1}+\ln x_{i-1}) 
            \Bigr)
         \biggl(A_{ff}^{(0)}+\sum_{f'} 
\Bigl(\max_z F^{(0)}_{f'f}(z)+M_{f'f}\Bigr)\biggr),
\label{phibarbintLOccfm}
\end{align}
where
\begin{align}
   \rho_{LO}(t+\ln x_{i-1},u)
  &= \frac{1}{\pi} \int du\, \alpha_{LO}(t+\ln x_{i-1}-u)
=- \frac{2}{\beta_0} \ln (2G(t+\ln x_{i-1},u)).
\end{align}
In order to generate $t_i$ the function $\bar\zeta_{LO}(t_i+\ln x_{i-1})$ is
then inverted either analytically or as a test option also numerically.

The generation of the flavor index $f_i$ is based on the probability
$p_{f_i}$ identical
to eqs.\ (\ref{flavindex}) and (\ref{flavindex2})

The generation of the $z$-variable is based on the LO analogue of eq.\
(\ref{zdistrc})
\begin{align}
\label{zdistrcLO}
 dz_ip^{C'(2)}(z_i)=& dz_i  \frac{\alpha_{LO}(t_i+ \ln
x_{i-1}+\ln(1-z_i))}{\pi}
         \frac{\Theta_{(1-z_i)x_{i-1}>\lambda e^{-t_i}}}{1-z_i}
\notag
\\ &
    \biggl(
    \int_0^1 dz \frac{\alpha_{LO}(t_i+ \ln x_{i-1}+ \ln(1-z))}{\pi}
         \frac{\Theta_{(1-z)x_{i-1}>\lambda e^{-t_i}}}{1-z}
    \biggr)^{-1} 
\notag
\\ 
   =& du_i \partial_{u_i}\rho_{LO}(t_i+ \ln x_{i-1},u_i) 
         \Theta_{t_i+ \ln x_{i-1}-t_\lambda>u_i>0}
\notag
\\ &
	\Bigl(\rho_{LO}(t_i+ \ln x_{i-1},t_i+ \ln x_{i-1}-t_\lambda)
                  -\rho_{LO}(t_i+ \ln x_{i-1},0)\Bigr)^{-1}.
\end{align}
As in the case B', the whole NLO effect is introduced through the global weight.
The virtual part of this weight, constructed on the base of the exact Sudakov
form factor derived for the
previous algorithm, follows from eq.\ (\ref{deltac})
\begin{align}
\label{deltaLOccfm}
\Delta_{f}^{C'(2)}&(t_i,t_{i-1}|x_{i-1})=
   \Phi_f^{C'}(t_i,t_{i-1}|x_{i-1}) -\bar{\Phi}_f^{C'(2)}(t_i,t_{i-1}|x_{i-1})
\notag
\\
=&
\int_{t_i}^{t_{i-1}}dt\int_0^1 dz  \theta_{(1-z)x_{i-1}>\lambda e^{-t}}\sum_{f'}
     \biggl(
    \frac{\alpha_{NLO}(t+\ln x_{i-1}+\ln(1-z))}{2\pi}2z
   P_{f'f}^{R(0)}(z)
\notag
\\ &
 + \Bigl(\frac{\alpha_{NLO}(t+\ln x_{i-1}+\ln(1-z))}{2\pi}\Bigr)^2 2zP_{f'f}^{R(1)C'}(z)
 \notag
 \\&
 - \frac{\alpha_{LO}(t+\ln x_{i-1}+\ln(1-z))}{2\pi}2z
   \bar P_{f'f}^{R(0)}(z)
  \biggr),
\end{align}
leading to
\begin{align}
\label{delta2LOccfm}
\Delta_{f}^{C'(2)}&(t_i,t_{i-1}|x_{i-1})=
\notag
\\
=&
\int_{t_{i-1}+\ln x_{i-1}-t_\lambda}^{t_i+\ln x_{i-1}-t_\lambda}du e^{-u} 2z
\sum_{f'} \biggl( 
      -
   \frac{(J_1^{LO}(t_i+\ln x_{i-1},u)-J_1^{LO}(u+t_\lambda,u))}{2\pi}
   \bar P_{f'f}^{R(C')}(z)
\notag
\\ &    +
   \frac{(J_1(t_i+\ln x_{i-1},u)-J_1(u+t_\lambda,u))}{2\pi}
   P_{f'f}^{R(0)}(z) 
   +
   \frac{(J_2(t_i+\ln x_{i-1},u)-J_2(u+t_\lambda,u))}{4\pi^2}P_{f'f}^{R(1)C'}(z)
          \biggr)
\notag
\\ &+
 \theta_{t_{\lambda}-\ln  x_{i-1}<t_{i-1}} \int_0^{t_{i-1}+\ln x_{i-1}-t_\lambda}du e^{-u} 2z
 \notag
 \\
 &
\sum_{f'} \biggl( 
     -
   \frac{(J_1^{LO}(t_i+\ln x_{i-1},u)-J_1^{LO}(t_{i-1}+\ln x_{i-1},u))}{2\pi}
   \bar P_{f'f}^{R(C')}(z)
\notag
\\ &+
   \frac{(J_1(t_i+\ln x_{i-1},u)-J_1(t_{i-1}+\ln x_{i-1},u))}{2\pi}
   P_{f'f}^{R(0)}(z) 
   \notag
    \\
    &+
   \frac{(J_2(t_i+\ln x_{i-1},u)-J_2(t_{i-1}+\ln x_{i-1},u))}{4\pi^2} P_{f'f}^{R(1)C'}(z)
          \biggr),\;\;\;\; z=1-e^{-u}.
\end{align}
As a result, the
complete formula for the global weight
relevant for this algorithm reads
\begin{equation}
\label{wtcorrcLOccfm}
w^{(n)(C')(2)}
= e^{-\Delta_{f_{n}}^{C'(2)}(t,t_{n}|x_{n})}
  \prod_{i=1}^n 
  \left(
\frac{      \Keu^{R(C')}_{f_if_{i-1}}(t_{i},x_i,x_{i-1})}%
           {\bar{\Keu}^{R(C')(2)}_{f_if_{i-1}}(t_{i},x_i,x_{i-1})} 
  e^{-\Delta_{f_{i-1}}^{C'(2)}(t_{i},t_{i-1}|x_{i-1})}
  \right).
\end{equation}
\section{Numerical results}
In this Section we present numerical results obtained with the MC
program {\tt EvolFMC} version 2. The first version of this program has been presented in the
Ref.\ \cite{Jadach:2003bu} for the case of standard DGLAP LO evolution. Subsequently, the NLO
evolution has been added in Ref.\ \cite{GolecBiernat:2006xw} and the LO modified-DGLAP evolution in 
Ref.\ \cite{GolecBiernat:2007pu}.
The presented here {\tt EvolFMC v.2} includes three of the described above four algorithms
for NLO modified-DGLAP evolution (the auxiliary algorithm for the B' evolution is not implemented at the moment). In order to accommodate these new evolution schemes,
the overall structure of the program has been modified, see Ref.\
\cite{stjinprep} for details. As a result, it was important to perform a number
of technical comparisons of the new code in order to establish its technical
precisions. We very briefly describe these tests in the following subsection. 
Later on we present numerical results regarding the
actual new evolution schemes. Before showing the results we 
discuss the choice of the counter terms and we list the input
parameters.
\subsection{Removal of the double counting}
\label{sect:double}
As indicated in Section \ref{sect:general},
modification of the argument of the coupling constant in the evolution
kernels is equivalent to adding some higher order (i.e.\ NLO and higher) terms.
Therefore, one has to make sure that there is no double counting with the
NLO part of the kernel. In the implementation presented here we
follow the approach as
discussed in Ref.\ \cite{Roberts:1999gb}. Namely, we take the  expansion of the $\alpha_{NLO}(t)$ in the form
\begin{align}
\alpha_{NLO}(t+\ln\phi) =
 \alpha_{NLO}(t) -\frac{\beta_0}{2\pi}\alpha_{NLO}^2(t) \ln\phi
 +{\cal O}(1/t^3),
\end{align}
where $\phi$ represents any arbitrary change in the argument.
Then the extra term in the kernels is of the form
\begin{align}
-{\beta_0} \ln\phi
  \Bigl(\frac{\alpha_{NLO}(t)}{2\pi}\Bigr)^2 2zP_{f'f}^{R(0)}(z)
\simeq
-{\beta_0} \ln\phi
  \Bigl(\frac{\alpha_{NLO}(t+\ln\phi)}{2\pi}\Bigr)^2 2zP_{f'f}^{R(0)}(z) 
+{\cal O}(1/t^3),
\end{align}
and consequently the counter terms from eqs.\ (\ref{non-universal}) could be
defined as follows
\begin{align}
\label{ctrterm}
&
\Delta P_{f'f}^{R(1)B'}(z) =
  {\beta_0} \ln (1-z) P_{f'f}^{R(0)}(z), \;\;\;\hbox{for scheme B'},
\\
&
\bar\Delta P_{f'f}^{R(1)C'}(z) =
  {\beta_0} \bigl(\ln w +\ln (1-z)\bigr) P_{f'f}^{R(0)}(z), \;\; z= x/w,
    \;\;\;\hbox{for scheme C'. } 
\label{cprim-ct}
\end{align}
However, a more detailed inspection of eq. (\ref{cprim-ct}) reveals that this
formula {\em over-subtracts} the double counting.
Namely, in the DGLAP evolution the kernels are functions of ``local''
$z$-variables only. In eq.\ (\ref{cprim-ct}) there is a corresponding $\ln (1-z)$
term. The other term, the $\ln w$, interconnects two emissions, and as such
it is of genuinely beyond-DGLAP origin. It means that it is absent in the DGLAP
kernel and there is no reason to remove it. As a result, the counter term
in the C' case becomes identical to the counter term in the B' case. 
In the following we will present results for both variants in the C' case, with
the understanding that the preferred choice is the 
\begin{align}
\label{universalCT}
&
\Delta P_{f'f}^{R(1)B'}(z) =
\Delta P_{f'f}^{R(1)C'}(z) =
  {\beta_0} \ln (1-z) P_{f'f}^{R(0)}(z), 
\end{align}
for both schemes B' and C'.
\subsection{Input parameters}
\label{sect:input}
The set-up of the {\tt EvolFMC} code is the same for all the presented results.
We use the gluon ($G$) and quark singlet
($Q$) PDFs with three massless quarks
\begin{equation}
D_Q =\sum_i\Bigl( D_{q_i} + D_{\bar{q}_i} \Bigr).
\end{equation}
%
As the initial distributions of the evolution we take
\begin{equation}
\label{inicond}
\begin{split}
D^0_G(x)&=1.908\cdot x^{-1.2}(1-x)^{5.0}, \\
D^0_{sea}(x)&=  0.6733\cdot x^{-1.2}(1-x)^{7.0}, \\
D^0_{u_{val}}(x)&=  2.187\cdot x^{-0.5}(1-x)^{3.0}, \\
D^0_{d_{val}}(x)&=  1.230\cdot x^{-0.5}(1-x)^{4.0}
\end{split}
\end{equation}
and
\begin{equation}
\label{inicond-quarks}
\begin{split}
D^0_{u}(x)&=  D^0_{u_{val}}(x)+\frac{1}{6} D^0_{sea}(x),\\
D^0_{d}(x)&=  D^0_{d_{val}}(x)+\frac{1}{6} D^0_{sea}(x),\\
D^0_{s}(x)&= 
D^0_{\bar{u}}(x)= 
D^0_{\bar{d}}(x)= 
D^0_{\bar{s}}(x)= \frac{1}{6} D^0_{sea}(x),\\
D^0_Q(x) &= D^0_{sea}(x)+D^0_{u_{val}}(x)+D^0_{d_{val}}(x).
\end{split}
\end{equation}
The QCD constant $\Lambda_{0}=0.2457$, the cut-off $\lambda=1$ GeV, $N_f=3$ and the dummy parameter
$\eta=0.1$. 
\subsection{Technical tests}
We have performed three different sets of the technical comparisons of the
code {\tt EvolFMC v.2}: 
\begin{enumerate}
\item
 With the semianalytical code {\tt APCheb40} \cite{APCheb40,GolecBiernat:2007xv}
based on the expansion in Chebyshev polynomials.
\item
 With the previous version of the {\tt EvolFMC} code: the {\tt EvolFMC v.1}, which was extensively tested in the past. 
\item
 Between different algorithms within the {\tt EvolFMC v.2}. It is the most important test of the new NLO modified-DGLAP evolutions, which are not available in any other code. 
\end{enumerate}
The overall conclusion of all the tests is that the technical precision of the program {\tt EvolFMC v.2} is at least $5\times 10^{-4}$ (half of a per mille). For the details of the tests we refer the reader to Ref.\ \cite{stjinprep}.
\subsection{Comparison of evolutions}
Having established the technical precision of the {\tt EvolFMC} code let us now proceed with the comparison of the various evolutions discussed earlier. Before getting into details let us remind the reader that in this paper for all of the results (i.e.\ for all of the evolutions) we use the same parameter setup: the same initial distributions at $Q=1$ GeV and the same $\Lambda_0$. In a more realistic study one should perform fits to the experimental data for each of the evolutions separately and then use different initial distributions for each evolution.

\begin{figure}[!ht]
 \centering
   \epsfig{file=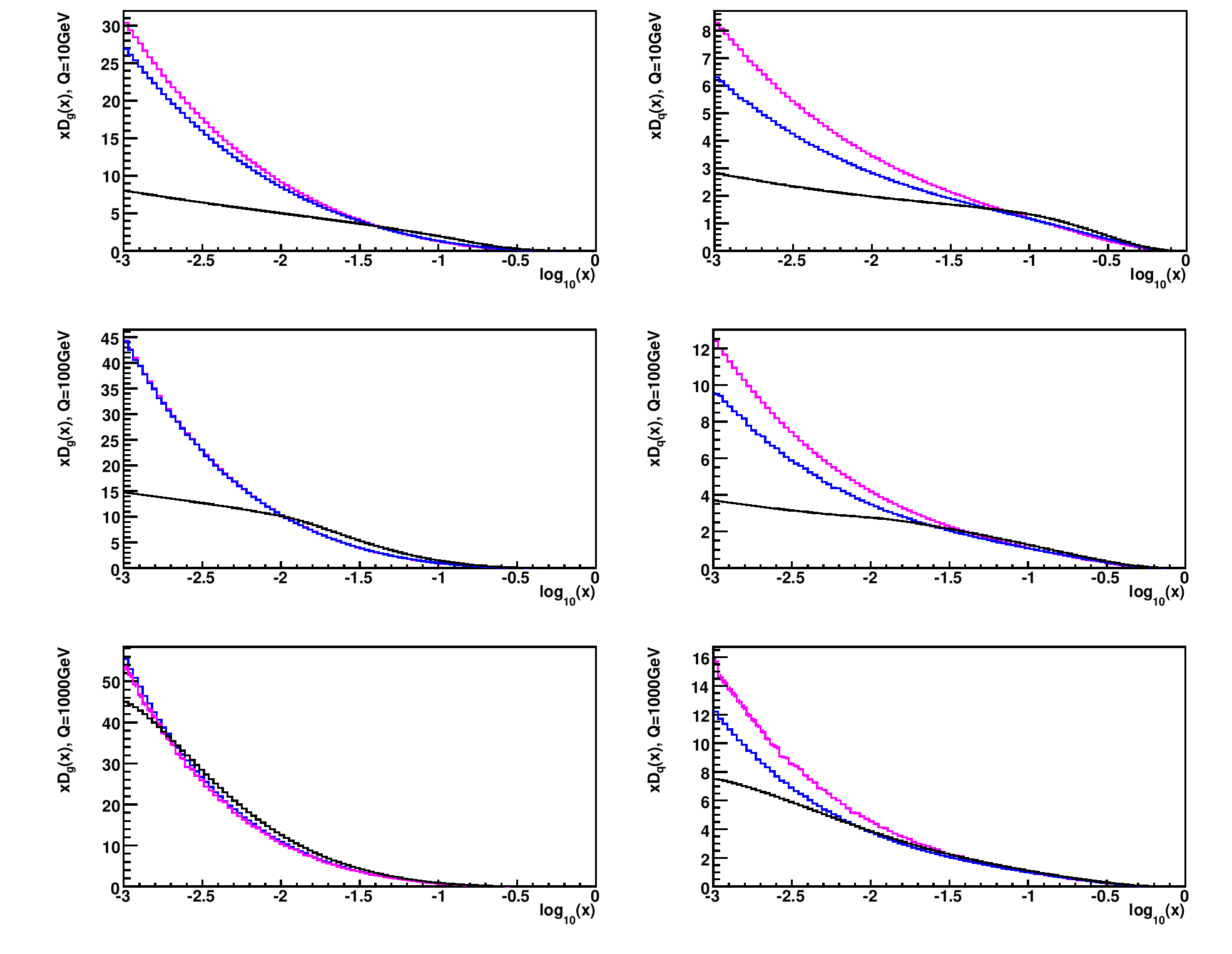,width=120mm,height=160mm,angle=90}
 \caption{\sf
Comparison of three NLO evolutions: DGLAP (magenta), B'-type (blue) and C'-type (black).
{\underline{Left frames}}: the gluon $xD_G(x)$ distr. 
{\underline{Right frames}}: the quark $xD_Q(x)$ distr.
{\underline{Top frames}}: the evolution up to 10 GeV.
{\underline{Central frames}}: the evolution up to 100 GeV.
{\underline{Bottom frames}}: the evolution up to 1000 GeV.
   }
 \label{fig:dglap_1mz_kt_nlo}
\end{figure}

We organize the numerical comparisons in the following way:
We begin by showing the "reference" result, i.e.\ the standard DGLAP in the LO and NLO approximations (Fig.\ \ref{fig:dglap_nlo_lo}). Next, we show the {\em new} results for the B'-type and C'-type evolutions. These two plots (Figs.\  \ref{fig:1mz_nlo_lo} and \ref{fig:cprim-bprim}) are the main numerical results of this paper, showing the NLO corrections in the modified DGLAP evolutions. The next two plots (Figs.\ \ref{fig:kt_nlo_lo} and \ref{fig:kt_nlo_lo_bezDC}) are of technical character
and show in a more detail certain aspects of the C'-type evolution. We conclude the section by comparing in a common plot all three evolutions in the NLO approximation (Fig.\ \ref{fig:dglap_1mz_kt_nlo}).
\subsubsection{NLO versus LO}
\begin{figure}[!ht]
 \centering
   \epsfig{file=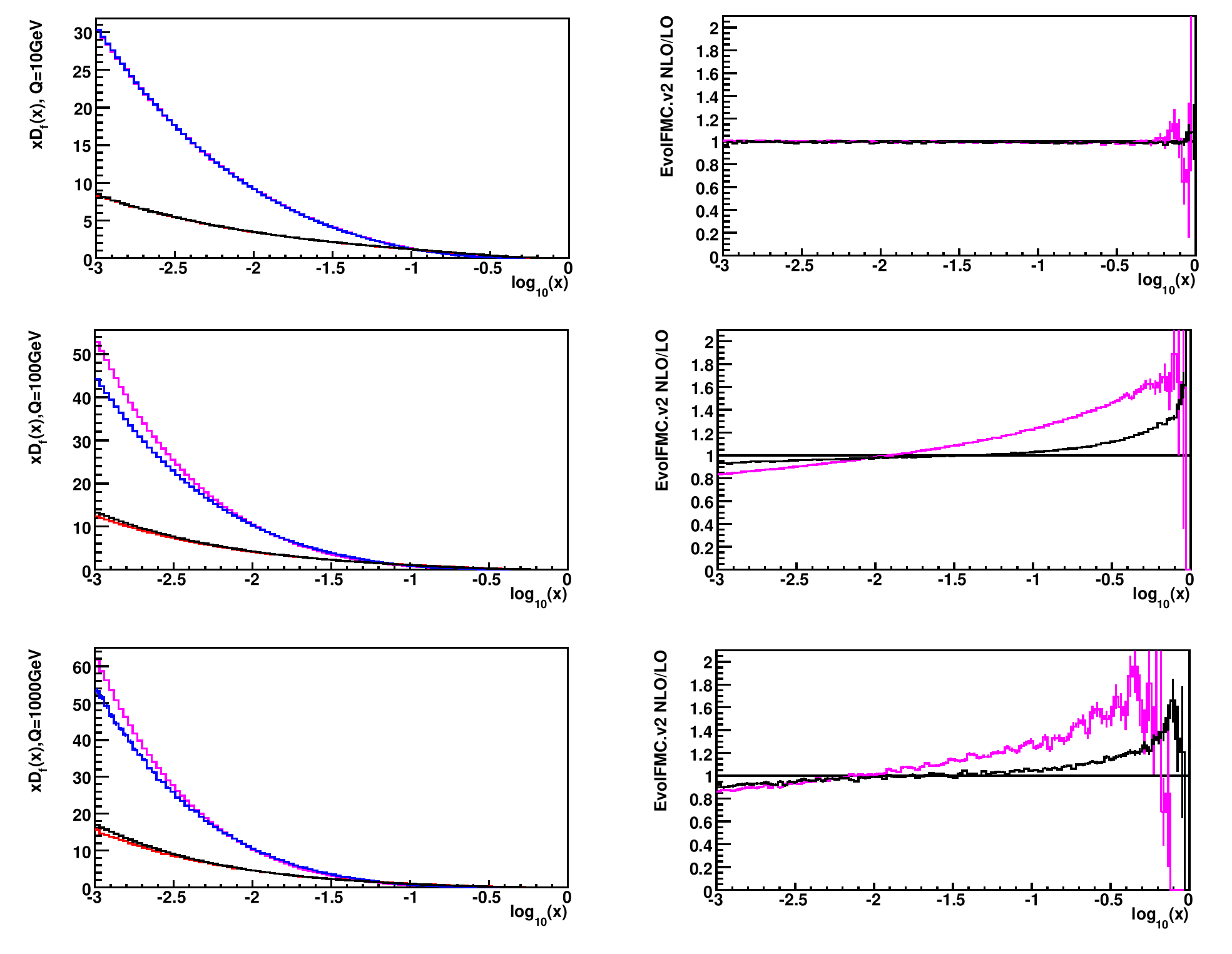,width=120mm,height=160mm,angle=90}
 \caption{\sf 
{\underline{Left frames}}: the DGLAP-type evolutions in the LO and NLO
approximations. 
Upper curves (LO: magenta and NLO: blue): the gluon $xD_G(x)$ distr.;
lower curves (LO: black and NLO: red): the quark $xD_Q(x)$ distr.
{\underline{Right frames}}:
the ratio of the NLO to LO distributions for the gluon  (magenta) and quark
(black)  
distributions.
{\underline{Top frames}}: the evolution up to 10 GeV.
{\underline{Bottom frames}}: the evolution up to 100 GeV.
    }
 \label{fig:dglap_nlo_lo}
\end{figure}
\begin{figure}[!ht]
 \centering
   \epsfig{file=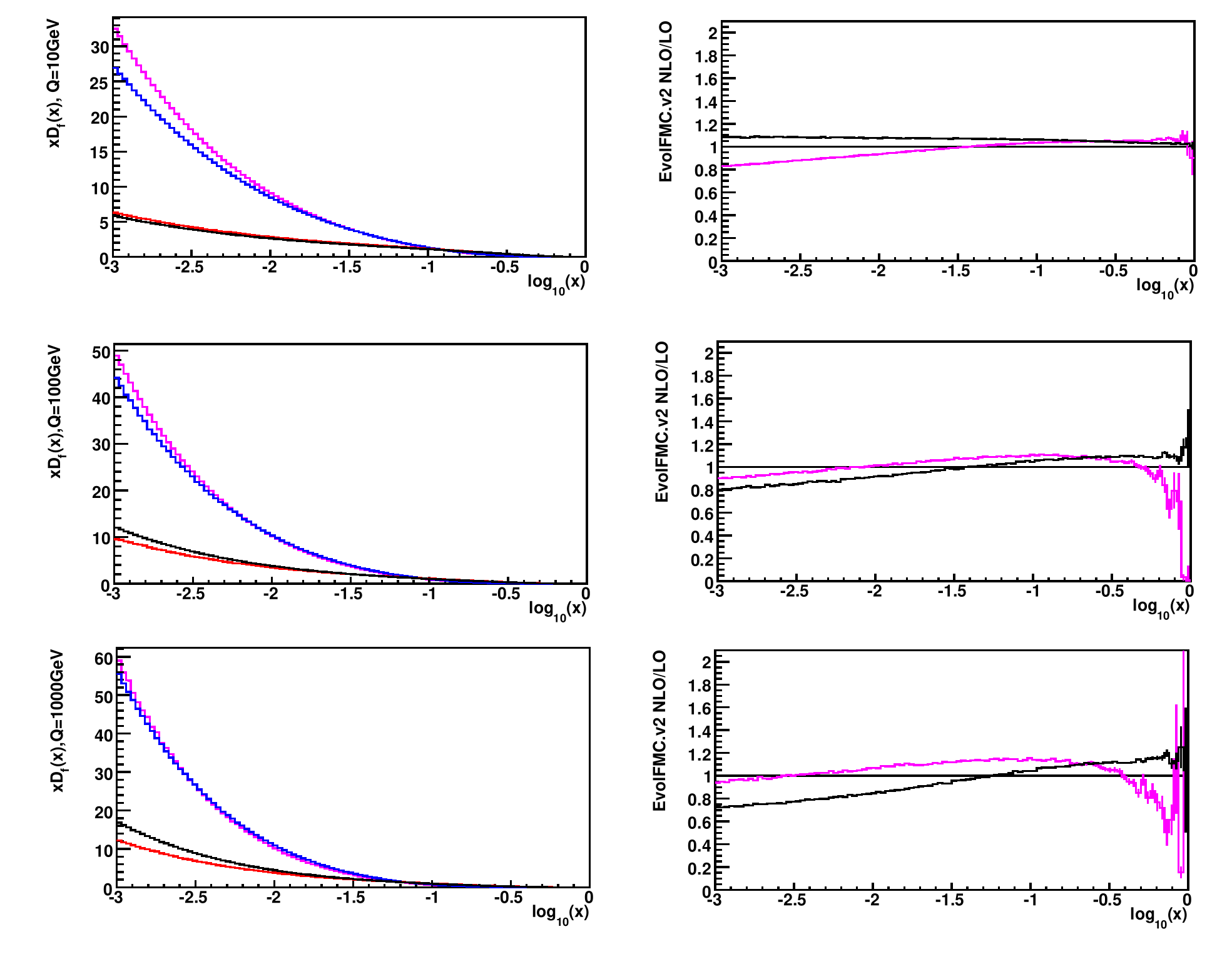,width=120mm,height=160mm,angle=90}
 \caption{\sf 
{\underline{Left frames}}: the modified-DGLAP B'-type evolutions in the LO and
NLO approximations. 
Upper curves (LO: magenta and NLO: blue): the gluon $xD_G(x)$ distr.;
lower curves (LO: black and NLO: red): the quark $xD_Q(x)$ distr.
{\underline{Right frames}}:
the ratio of the NLO to LO distributions for the gluon  (magenta) and quark
(black)  
distributions.
{\underline{Top frames}}: the evolution up to 10 GeV.
{\underline{Bottom frames}}: the evolution up to 100 GeV.
    }
 \label{fig:1mz_nlo_lo}
\end{figure}
\begin{figure}[!ht]
 \centering
   \epsfig{file=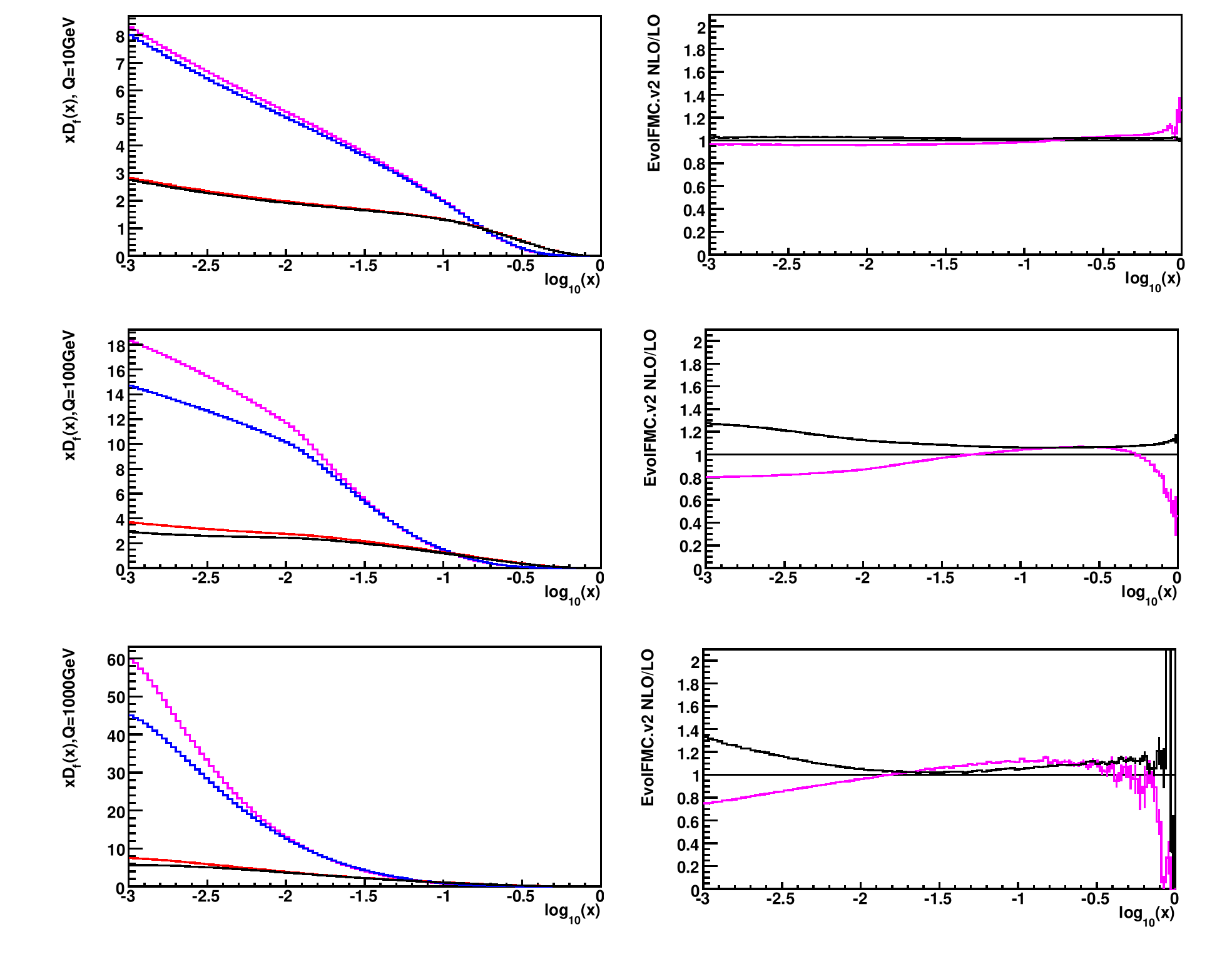,width=120mm,height=160mm,angle=90}
 \caption{\sf 
{\underline{Left frames}}: the modified-DGLAP C'-type evolutions in the LO and NLO approximations. 
Upper curves (LO: magenta and NLO: blue): the gluon $xD_G(x)$ distr.;
lower curves (LO: black and NLO: red): the quark $xD_Q(x)$ distr.
{\underline{Right frames}}:
the ratio of the NLO to LO distributions for the gluon  (magenta) and quark (black)  
distributions.
{\underline{Top frames}}: the evolution up to 10 GeV.
{\underline{Central frames}}: the evolution up to 100 GeV.
{\underline{Bottom frames}}: the evolution up to 1000 GeV.
    }
 \label{fig:cprim-bprim}
\end{figure}
\begin{figure}[!ht]
 \centering
   \epsfig{file=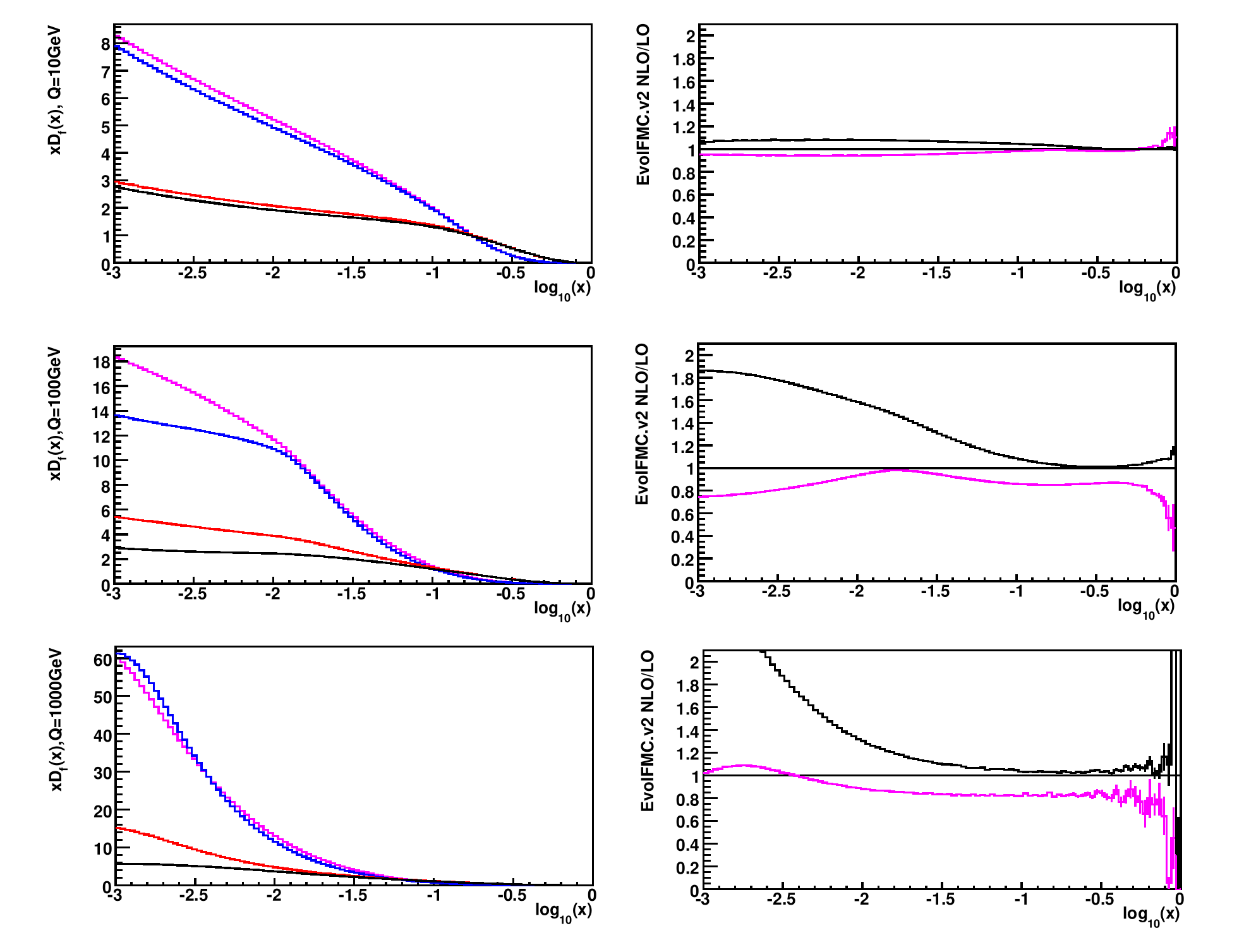,width=120mm,height=160mm,angle=90}
 \caption{\sf 
{\underline{Left frames}}: the modified-DGLAP C'-type evolutions in the LO and
NLO approximations with the {\em modified disfavoured (C'-type) counter term}. 
Upper curves (LO: magenta and NLO: blue): the gluon $xD_G(x)$ distr.;
lower curves (LO: black and NLO: red): the quark $xD_Q(x)$ distr.
{\underline{Right frames}}:
the ratio of the NLO to LO distributions for the gluon  (magenta) and quark
(black)  
distributions.
{\underline{Top frames}}: the evolution up to 10 GeV.
{\underline{Central frames}}: the evolution up to 100 GeV.
{\underline{Bottom frames}}: the evolution up to 1000 GeV.
    }
 \label{fig:kt_nlo_lo}
\end{figure}
\begin{figure}[!ht]
 \centering
   \epsfig{file=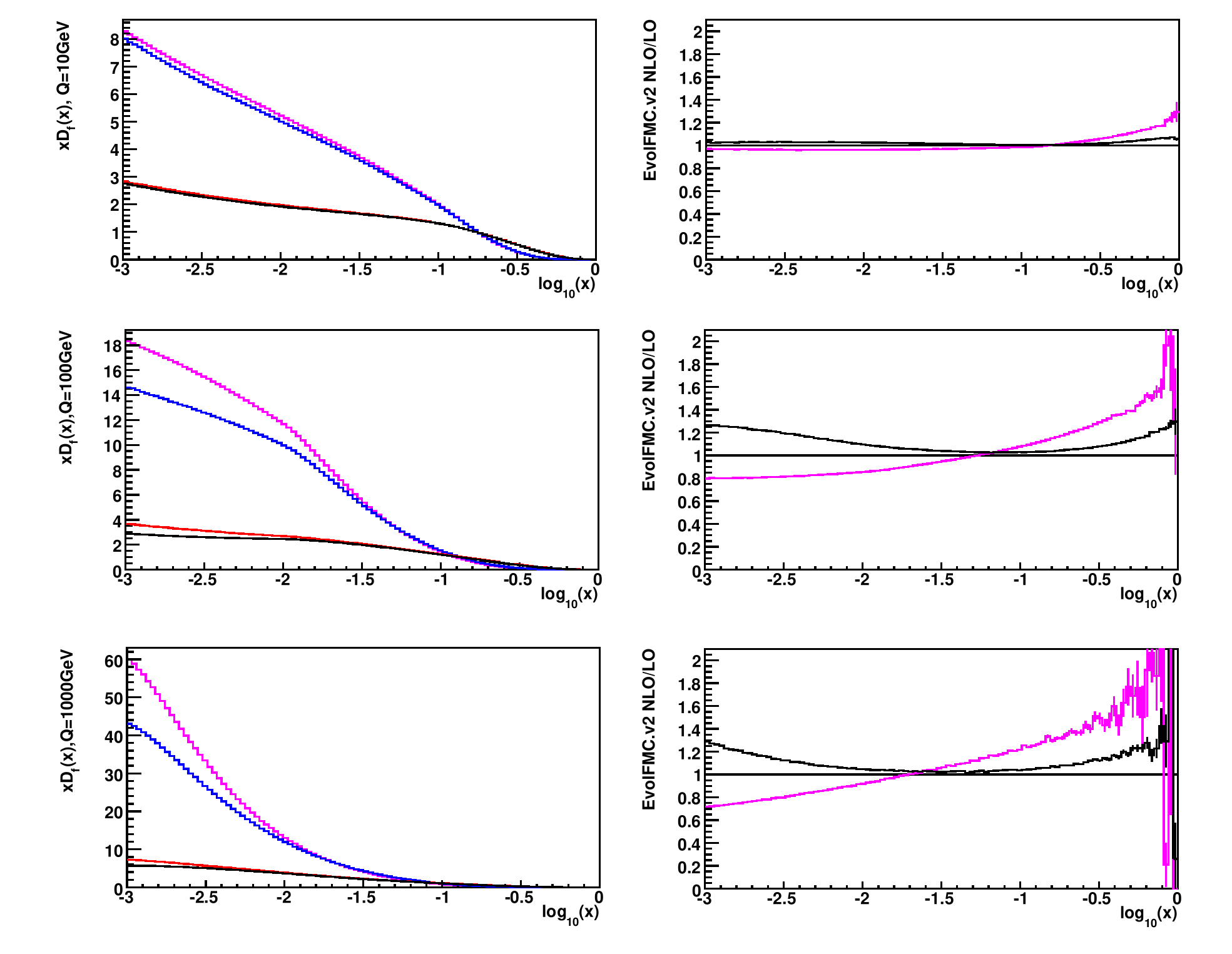,width=120mm,angle=90}
 \caption{\sf 
{\underline{Left frames}}: the modified-DGLAP C'-type evolutions in the LO and
NLO approximations without the counter term.
Upper curves (LO: magenta and NLO: blue): the gluon $xD_G(x)$ distr.;
lower curves (LO: black and NLO: red): the quark $xD_Q(x)$ distr.
{\underline{Right frames}}:
the ratio of the NLO to LO distributions for the gluon  (magenta) and quark
(black)  
distributions.
{\underline{Top frames}}: the evolution up to 10 GeV.
{\underline{Central frames}}: the evolution up to 100 GeV.
{\underline{Bottom frames}}: the evolution up to 1000 GeV.
    }
 \label{fig:kt_nlo_lo_bezDC}
\end{figure}
{\bf The "reference" DGLAP evolution}.
In the Fig.\ \ref{fig:dglap_nlo_lo} we show the case of the standard
DGLAP. We show these well known results because DGLAP
will serve us as a reference point in discussing the modified evolutions of 
the B'- and C'-type. We present the gluon and quark momentum distributions in LO and
NLO approximations as well as their ratios. Three evolution time limits are
shown: 10, 100 and 1000 GeV.
The characteristic feature of the plots is that DGLAP NLO corrections are
systematically bigger in the large $x$ region
and they show the tendency of diverging in the $x\to1$ limit. In the small $x$
region the NLO corrections are small. Results for the other evolutions will be
presented in a similar way. 
\\
{\bf The B'-type evolution} is shown in the Fig.\ \ref{fig:1mz_nlo_lo}.
It is one of the two main {\em new} numerical results of this paper.
As compared to the DGLAP case we can notice that the NLO corrections are a
bit bigger in the small $x$ region, of the size up to 20\%. In the large $x$
region, on the contrary, the corrections are much smaller and showing less
divergent behavior. This is in agreement with the general principle of
discussed here modifications of the DGLAP equation. These modifications 
are supposed to improve the description of the emission of soft partons
\cite{Amati:1980ch,Sterman:1986aj,Ermolaev:2008dq} and it is 
the $x\to 1$ limit which corresponds to the limit of only soft emissions. 
\\
{\bf The C'-type evolution.}
In the Fig.\ \ref{fig:kt_nlo_lo} we compare the LO and NLO evolutions
in the case of modified-DGLAP C'-type.
We consider this figure as the most important {\em new} numerical result of this paper.
We show the gluon and quark momentum distributions for two evolution time limits: 
10, 100 and 1000 GeV. 
We see that at 100 and 1000 GeV the NLO corrections in the small-$x$ region are
somewhat bigger than in the case B', reaching even 30\%. In the large $x$ region
the NLO corrections seem to be even milder and slightly less divergent than in
the B' case, showing the same improvement over DGLAP.
For the shorter time of 10 GeV the effects are much 
smaller. In fact, in this evolution type, due to the cut-off ${k_T>\lambda}$,
much less of the evolution happens before 10 GeV, and both the LO and NLO curves
are close to the initial condition as well as to each other.
Note
that in the case of both DGLAP and B', the evolution at 10 GeV is
already well developed. In order to reduce the evolution and get closer to the
initial condition, in the  DGLAP and B' cases the evolution time must be much shorter, 
below 2 GeV at least.
\\
{\bf Technical details related to C'-type evolution:}
\\
{\em The C'-type evolution with the disfavoured counter term $\bar\Delta P$ from eq.\
(\ref{cprim-ct}).} For the sake of comparison, in the Fig.\
\ref{fig:kt_nlo_lo}, we present also the other, disfavored, choice of the counter
term in the C' evolution, given in eq.\ (\ref{cprim-ct}). The plots clearly
confirm that the NLO corrections are much bigger, and, in addition, strongly
divergent in the small $x$ limit. 
\\
{\em The C'-type evolution with {\em no} counter term.}
As the last exercise, in Fig.\ \ref{fig:kt_nlo_lo_bezDC}, we completely
switched off the counter term in the C'-type evolution.
This way we try to understand better what actually causes the changes in
the shape of the NLO corrections in the modified schemes: is it the genuine
change of the argument of the coupling constant or rather the cut-off
$\lambda$?
In a crude approximation the counter term can be regarded as an estimate of the
size of the pure effect of shifting the argument in the coupling constant. 
As one can see, the shapes in the
right hand side plots in Fig.\ \ref{fig:kt_nlo_lo_bezDC} have changed
significantly in the large $x$ region, and have remained similar in the small
$x$ region, as compared to the complete C'
plots from Fig.\ \ref{fig:cprim-bprim}. This demonstrates that indeed it is the
change of the argument that drives the effect in the region of soft emission.
On the other hand, from the comparison of Figs.\ \ref{fig:cprim-bprim} and
\ref{fig:kt_nlo_lo_bezDC} to the DGLAP evolution, Fig.\ \ref{fig:dglap_nlo_lo},
we inferr that the higher order terms, resumed in the coupling
constant, contribute as much as the counter term.
\\
{\em Remark on the small $x$ limit of the C'-evolution.}
As we have mentioned, C' evolution is motivated by the CCFM equation. This equation
treats the $x\to 0$ limit in a better way than the DGLAP equation, in a sense
interpolating between DGLAP and BFKL equations. However, apart from the
modifications in the coupling constant and the angular ordering, which we have
incorporated in the C' scheme, there is one more ingredient missing in C' --
the 'non-Sudakov' form factor. This non-local form factor strongly
modifies the small $x$ evolution. It depends on the transverse momenta of all
the emitted partons. It is in principle not difficult to include such an effect
into the Monte Carlo evolution of the C'-type by means of the rejection
mechanism, provided the transverse momenta are properly generated within the
cascade, and we plan to do it in the future \cite{stjinprep}.

Let us summarize the comparisons:
1. The modifications of the argument of the coupling constant decrease the size
of the NLO corrections and reduce the divergences in the large $x$ region.
2. The NLO corrections in schemes B' and C' are significant, up to 30\% in
small $x$ region and must be included in the MC parton showers for the LHC.
3. One must remember, however, that these comparisons are to some degree
artificial because, as discussed earlier, we use the same input PDF
distributions for all evolution types. In a more complete study each of the evolutions should be separately fitted to the experimental data and obtained this way different input PDF distributions should be used.

\subsubsection{Different evolutions}
As the last exercise we compare the three analyzed types of the evolution.
In the Fig.\ \ref{fig:dglap_1mz_kt_nlo} we show simultaneously all three evolutions in the NLO approximation. It is clearly visible that the difference between DGLAP and B'-type evolution (with $\alpha(Q(1-z))$) is rather small and of the quantitative form. On the contrary, the C'-type evolution (with $\alpha(k_T)$) looks very different, both in the shape
and in the magnitude. There is a visible flattening of the C'-type distributions around the value of  
$x\sim \lambda/Q$ caused by the cut-off $\lambda$. 
This follows from the condition $x_{i-1}(1-z_i)\geq \lambda/Q$ which stops any
cascade as soon as the $x_i$ variable falls below the $\lambda/Q$ threshold.
In particular no evolution at all can develop for any $x_0\leq \lambda/Q$.

\section{Summary and outlook}
In this paper we have presented a series of Markovian MC algorithms that solve the QCD evolutions of the modified-DGLAP type in the NLO approximation. One of the two discussed modifications of the DGLAP evolution is of high practical importance. In this evolution, as the argument of the coupling constant the transverse momentum of the emitted parton is 
used, and the evolution time is identified with the rapidity variable. Such a modification 
is known to describe better the emission of soft partons. It can serve as
a first step towards incorporating the complete CCFM effects into the 
evolution as well.
In this paper we have called this scenario the C'-type evolution. The other
scenario, called throughout the paper the  B'-type evolution, uses $Q(1-z)$ as
the argument of the coupling constant, modelling the "one-loop" CCFM evolution
equation. It is one of many possible $z$-dependent modifications of the argument
discussed in the literature. Proper counter terms have been added to the
evolution kernels in order to remove double counting at the NLO level.

The algorithms have been 
implemented into the new version of the MC program {\tt EvolFMC} and extensively tested by comparisons with the semianalytical code {\tt APCheb40}, with the previous version of {\tt EvolFMC} and by comparisons of independent algorithms within the new version of  
{\tt EvolFMC} itself. As the overall conclusion of the tests we claim the technical precision
of the {\tt EvolFMC} to be at least $5\times 10^{-4}$.

The comparison of the modified-DGLAP evolutions at both LO and NLO level shows
that the NLO corrections are in general smaller in the modified evolutions and
more importantly, the divergent behavior of the NLO corrections in the large
$x$ region is limited, as expected. The only exception is the very low $x$
region where the NLO corrections in the modified schemes are larger. This is
however the region where the DGLAP equation becomes less accurate anyway.

Quantitatively, the NLO corrections are 
relatively modest: in the B' case they are
small, of the order of up to 20\% of the LO terms.
For the $k_T$-dependent evolution (C') they are, in most of the parameter space,
of the order of 10\%, but in some limited regions of small $x$ they grow up to 30\%.
These results, on the one hand, show that the NLO contributions are numerically
significant and should be taken into account in the construction of the parton
shower MC for the LHC experiments. This is especially important in the case of the
physically well motivated $k_T$-dependent evolution. On the other hand, the
convergence of the QCD perturbative expansion looks reasonably well, better
than in the DGLAP case.

The main limitations of the presented in this paper MC algorithms are:
missing effects due to the non-zero masses of the quarks in all of the discussed
types of evolution and lack of the dedicated fits to the data for the
modified-DGLAP type evolutions. Another interesting development line of the
modified-DGLAP type evolutions would be to include a non-perturbative
parametrization of the behavior of the coupling constant below the Landau pole.
We hope to address some of these issues in the future.

\section{Acknowledgments}
The authors would like to thank S. Jadach, K. Golec-Biernat and Z. Was for numerous discussions and comments.

\providecommand{\href}[2]{#2}
\end{document}